\documentclass[11pt,onecolumn,english,preprintnumbers,amsmath,amssymb,floatfix,nofootinbib]{revtex4}

\usepackage[T1]{fontenc}
\usepackage[latin9]{inputenc}
\usepackage{color}
\usepackage{array}
\usepackage{amstext}
\usepackage{graphicx}
\usepackage{esint}
\usepackage{rotating}
\usepackage{subfigure}

\usepackage{array}
\newcolumntype{L}[1]{>{\raggedright\let\newline\\\arraybackslash\hspace{0pt}}m{#1}}
\newcolumntype{C}[1]{>{\centering\let\newline\\\arraybackslash\hspace{0pt}}m{#1}}
\newcolumntype{R}[1]{>{\raggedleft\let\newline\\\arraybackslash\hspace{0pt}}m{#1}}
\usepackage[font=small,labelfont=bf]{caption}
\usepackage{xcolor}
\usepackage[colorlinks = true,
linkcolor = magenta,
urlcolor  = blue,
citecolor = red,
anchorcolor = blue]{hyperref}

\makeatletter

\@ifundefined{textcolor}{}
{%
 \definecolor{BLACK}{gray}{0}
 \definecolor{WHITE}{gray}{1}
 \definecolor{RED}{rgb}{1,0,0}
 \definecolor{GREEN}{rgb}{0,1,0}
 \definecolor{BLUE}{rgb}{0,0,1}
 \definecolor{CYAN}{cmyk}{1,0,0,0}
 \definecolor{MAGENTA}{cmyk}{0,1,0,0}
 \definecolor{YELLOW}{cmyk}{0,0,1,0}
 }

\@ifundefined{definecolor}
 {\usepackage{color}}{}
\@ifundefined{definecolor}
 {\usepackage{color}}{}
\makeatother
\usepackage{babel}

\def\Re{{\cal R \mskip-4mu \lower.1ex \hbox{\it e}\,}}
\def\Im{{\cal I \mskip-5mu \lower.1ex \hbox{\it m}\,}}

\def\tev{\,{\ifmmode\mathrm {TeV}\else TeV\fi}}
\def\gev{\,{\ifmmode\mathrm {GeV}\else GeV\fi}}
\def\mev{\,{\ifmmode\mathrm {MeV}\else MeV\fi}}
\def\to{\rightarrow}

\begin{document}


\title { Study of Higgs Effective Couplings at Electron-Proton Colliders  }
	
\author { Hoda Hesari$^{1}$  }
\email{h.hesari@ipm.ir}

\author { Hamzeh Khanpour$^{2,1}$  }
\email{ Hamzeh.Khanpour@mail.ipm.ir }

\author{ Mojtaba Mohammadi Najafabadi$^{1}$  }
\email{mojtaba@ipm.ir}

\affiliation {
$^{(1)}$School of Particles and Accelerators, Institute for Research in Fundamental Sciences (IPM), P.O.Box 19395-5531, Tehran, Iran   \\
$^{(2)}$Department of Physics, University of Science and Technology of Mazandaran, P.O.Box 48518-78195, Behshahr, Iran   }

\date{\today}

%
\begin{abstract}\label{abstract}

We perform a search for beyond the standard model dimension-six operators
relevant to the Higgs boson at the Large Hadron Electron Collider (LHeC) and the Future Circular Hadron Electron Collider (FCC-he).
With a large amount of data (few ab$^{-1}$) and collisions at TeV scale, both LHeC and FCC-he provide  excellent opportunities
to search for the BSM effects.
The study is done through the process $e^-p \to h j \nu_e$ where the Higgs boson decays into a pair of $b \bar{b}$
and we consider the main sources of background processes including a realistic simulation of detector effects.
For the FCC-he case, in some signal scenarios to obtain an efficient event reconstruction and to have a good background rejection, 
jet substructure techniques are employed to reconstruct the boosted Higgs boson in the final state.
In order to assess the sensitivity  to the dimension-six operators, a shape analysis on the differential cross sections is performed.
Stringent bounds are found on the Wilson coefficients
of dimension-six operators with the integrated luminosities of 1 ab$^{-1}$ and 10 ab$^{-1}$ which in some cases show improvements
with respect to the high-luminosity LHC results.

\end{abstract}
%


\maketitle

\newpage

\tableofcontents{}

%
\section{Introduction}\label{sec:intro}

So far, the Standard Model (SM) of particle physics has been found to
be a successful theory describing nature up to the scale of electroweak. However,  there are
reasons to believe that the SM is not the ultimate theory of particle physics at the TeV scale.
The Higgs boson discovery by the LHC experiments~\cite{Aad:2012tfa, Chatrchyan:2012xdj} has been a milestone
in understanding the mechanism of electroweak symmetry breaking (EWSB).
After that, one of the main goals would be to precisely  measure the Higgs boson properties
that would provide the possibility of searching for new physics effects beyond the SM.

Given the absence of any signature of new physics in the present data,
one can parametrize the effects of beyond the SM in an effective field theory
(EFT) expansion.
This approach is a powerful tool which parameterizes possible new physics effects via a systematic expansion in a series of higher-dimensional
operators composed of SM  fields~\cite{Buchmuller:1985jz,Grzadkowski:2010es}.
The operators are composed of all possible combinations of SM fields  respecting the
$SU(3)_c \times SU(2)_{\mathrm L} \times U(1)_{\mathrm Y}$ gauge symmetries and Lorentz invariance.
In the EFT approach, potential deviations from the SM could be described using the following Lagrangian:
\begin{equation}\label{SM-EFT}
\mathcal{L}_{\rm SM \, EFT} = \mathcal{L}_{\rm SM} + \sum_i \frac{c_i}{\Lambda^{2}} \mathcal{O}_i + {\rm h.c.} 	\, ,
\end{equation}
where the $\mathcal{O}_i$ is the $i$th dimension-six operator, $\Lambda$ is the scale at which new physics is expected to appear
and $c_{i}'$s are arbitrary Wilson coefficients. These dimension-six operators have been listed and studied in  Refs.~\cite{Buchmuller:1985jz,AguilarSaavedra:2008zc,Grzadkowski:2010es,Arzt:1994gp}. There have been a lot of studies to probe these operators and so far a lot of attention has been paid to constrain these operators that can be found in Refs.~\cite{Hartmann:2016pil,Kuday:2017vsh,Kilian:2017nio,Ellis:2017kfi,Fichet:2016iuo,Sigismondi:2012sp,Arbey:2016kqi,Amar:2014fpa,Banerjee:2015bla,Craig:2015wwr,Corbett:2015ksa,Ellis:2014jta,Berthier:2015gja,Englert:2015hrx,Ellis:2014dva,Khanpour:2017cfq,Khanpour:2017inb,Buckley:2015lku,Buckley:2015nca,Denizli:2017pyu,Barklow:2017suo,Murphy:2017omb,Jana:2017hqg,Dedes:2017zog,Dedes:2018seb}.

The aim of this study is to explore Wilson coefficients of dimension-six operators, as described in Refs.~\cite{Contino:2013kra,Artoisenet:2013puc,Alloul:2013naa},
contributing to the Higgs production
in association with a jet and a neutrino at the LHeC and FCC-he~\cite{Bruening:2013bga,AbelleiraFernandez:2012cc,AbelleiraFernandez:2012ty,AbelleiraFernandez:2012ni}.
The LHeC is a proposed deep inelastic electron-nucleon scattering (DIS)
machine which has been designed to collide electrons with an energy from 60 GeV to possibly 140 GeV,
with protons with an energy of 7 TeV.
The future circular collider (FCC) has the option of colliding electrons to protons
with the electron energy $E_e$ = 60 GeV and the proton energy of $E_p$ = 50 TeV.
The inclusive Higgs boson production cross section at the high-energy FCC-he is expected to be about five times
larger than at the future proposed high-energy and high-luminosity electron-positron collider TLEP/FCC-ee~\cite{Kumar:2015kca}.
In comparison with the LHC or FCC-hh, the LHeC or FCC-he have the advantages of providing a clean environment
with small background contributions from QCD strong interactions.
Furthermore, no effects of pile-up and multiple interactions exist in these machines
and they are able to provide precise measurements of the proton structure, electroweak and strong interactions.

The remaining part of this paper is structured as follows:  In Section~\ref{sec:Framework}, we present the theoretical framework of our
analysis by recalling the relevant aspects of the effective SM in which dimension-six operators are considered.
In this section, we review the higher-dimensional operators and highlight the operators contributing to
the Higgs boson production processes at the LHeC and FCC-he.
Section~\ref{sec:Simulation} describes the details of our analysis including the
simulation tools and analysis strategy for both LHeC and FCC-he.
We will explain the event selection criteria and statistical method by which we obtain the constraints on
the Wilson coefficients. The analysis strategy for the LHeC collider and its sensitivity to the dimension-six operators have been presented
in Section~\ref{sec:LHeC}. Section~\ref{sec:FCC-he} is dedicated to present the sensitivity of the FCC-he collider to the related Wilson coefficients in the $h j \nu_{e}$ process.
Our results and the constraints on the Wilson coefficients are given in Section~\ref{sec:results}. Comparisons with the 
LHC bounds are made also in this section.
Finally, Section~\ref{sec:Discussion} presents the summary and conclusions.

%
\section{Theoretical framework}\label{sec:Framework}

The SM effective Lagrangian can be obtained by including higher-dimensional operators
that takes into account the new physics effects  beyond the SM which may appear at the energy scale
much larger than  the SM energy scale.  Under the assumption of baryon and lepton number
conservation and keeping only dimension-six operators,
the most general  $SU(3)_{\rm C} \times SU(2)_{\rm L}\times U(1)_{\rm Y}$ gauge invariant Lagrangian
can be constructed from the SM fields.
We concentrate on the dimension-six interactions of the Higgs boson, fermions, and the electroweak
gauge bosons in the strongly interacting light Higgs (SILH) basis conventions which can be written 
as~\cite{Contino:2013kra,Pomarol:2013zra,Alloul:2013naa}:
\begin{equation}\label{eq:effLag}
{\cal L}_{\rm eff} = {\cal L}_{\rm SM}  + \sum_i {\bar c_i} {\cal O}_i \equiv {\cal L}_{\rm SM} + \Delta {\cal L}_{F_1} +  \Delta {\cal L}_{F_2} + \Delta {\cal L}_{\rm SILH} \,,
\end{equation}
where $\bar{c}_{i}$ coefficients are dimensionless Wilson coefficients, and ${\cal O}_i$ are dimension-six operators made up of SM fields.
The first term in the effective Lagrangian of Eq.~\eqref{eq:effLag} is the SM Lagrangian, ${\cal L}_{\rm SM}$.
The second term $\Delta {\cal L}_{F_1}$ in Eq.~\eqref{eq:effLag} addresses the interactions between two Higgs fields and a pair of quarks or leptons. This
term has the following form:
\begin{equation}\label{eq:silh2}
\begin{split}
\Delta {\cal L}_{F_1} =
\, & \frac{i \bar c_{HQ}} {v^2} \left(\bar q_L \gamma^\mu q_L\right) \big( H^\dagger {\overleftrightarrow D}_\mu H \big)
 + \frac{i\bar c^\prime_{HQ}} {v^2} \left(\bar q_L \gamma^\mu \sigma^i q_L\right) \big(H^\dagger\sigma^i {\overleftrightarrow D}_\mu H \big)   \\
& + \frac{i\bar c_{Hu}} {v^2} \left(\bar u_R \gamma^\mu u_R\right) \big( H^\dagger {\overleftrightarrow D}_\mu H \big)    \\
& + \frac{i\bar c_{Hd}} {v^2} \left(\bar d_R \gamma^\mu d_R\right) \big( H^\dagger {\overleftrightarrow D}_\mu H \big)
 + \bigg[\frac{i \bar c_{Hud}} {v^2} \left(\bar u_R \gamma^\mu d_R\right) \big( H^{c\, \dagger} {\overleftrightarrow D}_\mu H\big) +{\it h.c.} \bigg]   \\
& + \frac{i\bar c_{HL}} {v^2} \left(\bar L_L \gamma^\mu L_L\right) \big( H^\dagger {\overleftrightarrow D}_\mu H \big)    \\
& + \frac{i\bar c^\prime_{HL}} {v^2} \left(\bar L_L \gamma^\mu \sigma^i L_L\right) \big(H^\dagger\sigma^i {\overleftrightarrow D}_\mu H \big)
 + \frac{i\bar c_{Hl}} {v^2} \left(\bar l_R \gamma^\mu l_R \right) \big( H^\dagger {\overleftrightarrow D}_\mu H \big) \,.
\end{split}
\end{equation}
The third term $\Delta {\cal L}_{F_2}$ of the effective Lagrangian in Eq.~\eqref{eq:effLag}
contains the interactions of a pair of quark or lepton, a Higgs field,  and a gauge boson. This term reads:
\begin{equation}\label{eq:silh3}
\begin{split}
\Delta {\cal L}_{F_2} =
\, & \frac{ \bar c_{uB}\,g^\prime} {m_W^2}\, y_u \, {\bar q}_L H^c \sigma^{\mu \nu} u_R \, B_{\mu \nu}
 + \frac{ \bar c_{uW}\,g}{m_W^2}\,  y_u \, {\bar q}_L  \sigma^i H^c \sigma^{\mu \nu} u_R \, W_{\mu \nu}^i    \\
& + \frac{ \bar c_{uG}\,g_S}{m_W^2}\, y_u \, {\bar q}_L H^c \sigma^{\mu\nu} \lambda^a u_R \, G_{\mu \nu}^a
\, + \frac{ \bar c_{dB}\,g^\prime } {m_W^2}\, y_d \, {\bar q}_L H \sigma^{\mu \nu} d_R \, B_{\mu \nu}         \\
& + \frac{ \bar c_{dW}\,g}{m_W^2}\, y_d \, {\bar q}_L \sigma^i H \sigma^{\mu \nu} d_R \, W_{\mu \nu}^i        \\
& + \frac{ \bar c_{dG}\,g_S}{m_W^2}\, y_d \, {\bar q}_L H \sigma^{\mu \nu}  \lambda^a d_R \, G_{\mu \nu}^a
\, + \frac{ \bar c_{lB}\,g^\prime } {m_W^2}\, y_l \, {\bar L}_L H \sigma^{\mu \nu}  l_R \, B_{\mu \nu}        \\
& + \frac{ \bar c_{lW}\,g}{m_W^2}\, y_l \, {\bar L}_L \sigma^i H \sigma^{\mu \nu} l_R \, W_{\mu \nu}^i + {\text h.c.} \,
\end{split}
\end{equation}
Finally, the last term of this Lagrangian corresponds to the Higgs field which is the part of a strongly interacting light Higgs sector (SILH).
The $\Delta {\cal L}_{\rm SILH}$ term can be expressed as:
\begin{equation}\label{eq:silh}
\begin{split}
\Delta {\cal L}_{\rm SILH} =
\, & \frac {\bar c_H} {2v^2} \, \partial^\mu \, \left( H^\dagger H \right) \partial_\mu \, \left( H^\dagger H \right)
 + \frac {\bar c_T} {2v^2} \left (H^\dagger {\overleftrightarrow {D^\mu}} H \right) \,\left(H^\dagger {\overleftrightarrow D}_\mu H \right)
 - \frac {\bar c_6 \, \lambda} {v^2} \left( H^\dagger H \right)^3        \\
& + \bigg[  \bigg[ \frac{\bar c_u} {v^2} \, y_{u}\, H^\dagger H\, {\bar q}_L H^c u_R + \frac{\bar c_d}{v^2}\, y_{d} \, H^\dagger H \, {\bar q}_L H d_R
 + \frac {\bar c_l} {v^2}\,y_{l}\, H^\dagger H\, {\bar L}_L H l_R \bigg] + {\text h.c.} \bigg]     \\
& + \frac {i\bar c_W\, g} {2m_W^2} \left( H^\dagger \sigma^i \overleftrightarrow {D^\mu} H \right )( D^\nu  W_{\mu \nu})^i    \\
& + \frac {i\bar c_B\, g{\prime}} {2m_W^2} \left(H^\dagger \overleftrightarrow {D^\mu} H \right)(\partial^\nu B_{\mu \nu})
 + \frac {i \bar c_{HW} \, g} {m_W^2}\, (D^\mu H)^\dagger \sigma^i (D^\nu H) W_{\mu \nu}^i   \\
& + \frac {i\bar c_{HB} \, g^{\prime} } {m_W^2} \, (D^\mu H)^\dagger (D^\nu H) B_{\mu \nu}
 + \frac {\bar c_\gamma \, {g {\prime}}^2} {m_W^2} \, H^\dagger H B_{\mu \nu} B^{\mu \nu}
 + \frac {\bar c_g \, g_S^2} {m_W^2} \, H^\dagger H G_{\mu \nu}^a G^{a\mu \nu}    \\
& + \frac {i \tilde c_{HW} \, g} {m_W^2} \, (D^\mu H)^\dagger \sigma^i (D^\nu H) {\tilde W}_{\mu \nu}^i
 +\frac {i\tilde c_{HB} \,g^{\prime} } {m_W^2} \, (D^\mu H)^\dagger (D^\nu H) {\tilde B}_{\mu \nu}   \\
&  +\frac {\tilde c_\gamma \,{g {\prime}}^2} {m_W^2} \, H^\dagger H B_{\mu \nu} {\tilde B}^{\mu \nu}
+\frac {\tilde c_g \, g_S^2} {m_W^2} \, H^\dagger H G_{\mu \nu}^a{\tilde G}^{a\mu\nu}     \\
\, &
+ \frac {\tilde c_{3W} \, g^3} {m_W^2} \, \epsilon^{i  jk} W_{\mu}^{i \,\nu} W_{\nu}^{j \, \rho} {\tilde W}_{\rho}^{k \,\mu}
 + \frac {\tilde c_{3G} \, g_S^3} {m_W^2} \,f^{abc} G_{\mu}^{a \,\nu} G_{\nu}^{b \,\rho} {\tilde G}_{\rho}^{c \,\mu} \,.
\end{split}
\end{equation}
where $\Phi$ is a weak doublet containing the Higgs boson field, and  $G^{\mu\nu}$,  $B^{\mu\nu}$,  $W^{\mu \nu}$ are the
strong and electroweak field strength tensors. $\Phi^\dag \overleftrightarrow{D}^\mu \Phi = \Phi^\dag (D^{\mu}\Phi) - (D^{\mu}\Phi)^{\dag}\Phi$
is the hermitian covariant derivative. In Eq.\ref{eq:silh},
 $\lambda$ is the Higgs boson quartic coupling and $v$ is the vacuum expectation value defined as $v = 1/(\sqrt{2}G_{F})^{1/2} = 246$ GeV.

\begin{figure}[!htbp]
	\begin{center}
		\includegraphics[width=7.0cm]{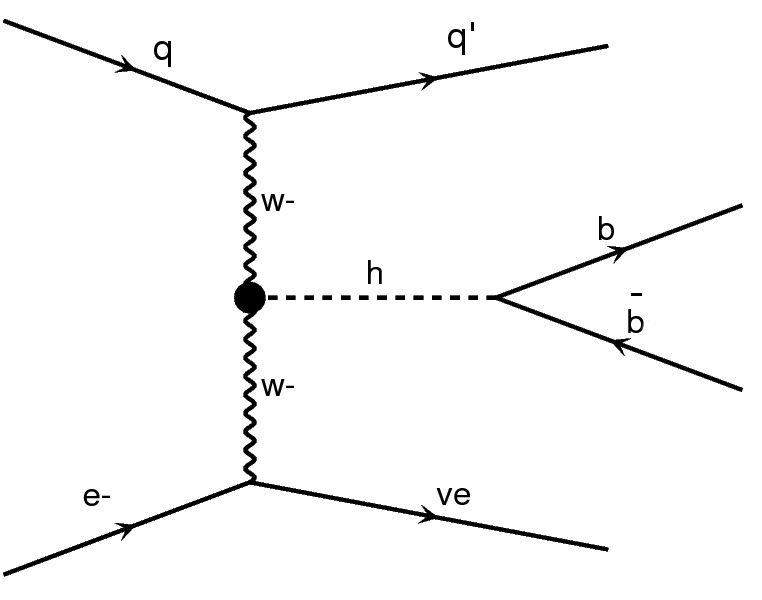}
		\hspace{5mm}
		\caption{ Leading order Feynman diagram Higgs boson production via $e^-p \to h j \nu_e$ processes. }
		\label{fig:higgs_diagram}
	\end{center}
\end{figure}

In the electron-proton colliders, the Higgs bosons are produced through two main  channels.
The Higgs boson can be produced either via charged current: $e^{-}q \rightarrow Hq'\nu_{e}$
or neutral current $e^{-}q\rightarrow eHq$. The leading order diagram
for the production of a Higgs boson in the electron-proton collisions for the charged current process
is depicted in Fig.~\ref{fig:higgs_diagram}. There are already several studies on different aspects of the Higgs boson production
via charged and neutral currents in the electron-proton collisions
~\cite{Ellis:1975ap,LoSecco:1976ii,Hioki:1983yz,Blumlein:1992eh,Han:2009pe,Biswal:2012mp,Sun:2016kek,Wang:2017pdg,Senol:2012fc}.
In Ref.~\cite{Han:2009pe}, it has been shown that the production cross section of the charged current process
is larger than  the neutral current process by  a factor of a around five for the energy of the incoming
electron 140 GeV and the proton energy of 7 TeV.
In this work, our focus is on the charged current production process due to its larger production cross section.
Also, this process has a clean signature as it comprises of a significant missing transverse energy and an energetic jet which tends to be forward.
The concentration of this analysis is on the Higgs boson decay into a pair of bottom quarks because of its large branching fraction.

\begin{figure}[!htbp]
	\centering
	\subfigure(a){\includegraphics[width=6.0cm]{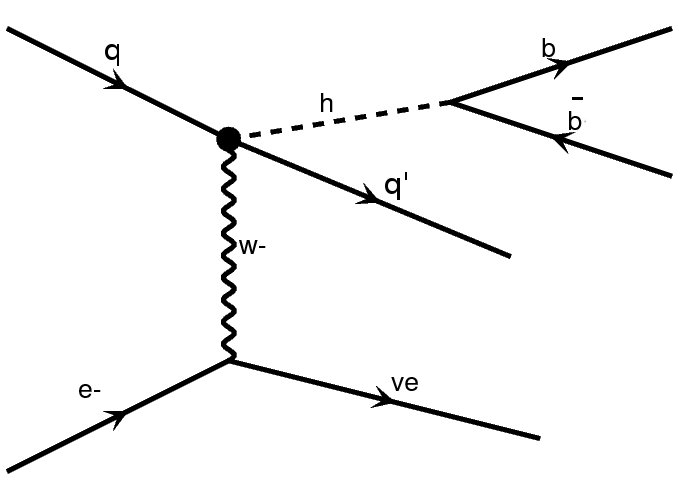}}
	\subfigure(b){\includegraphics[width=6.0cm]{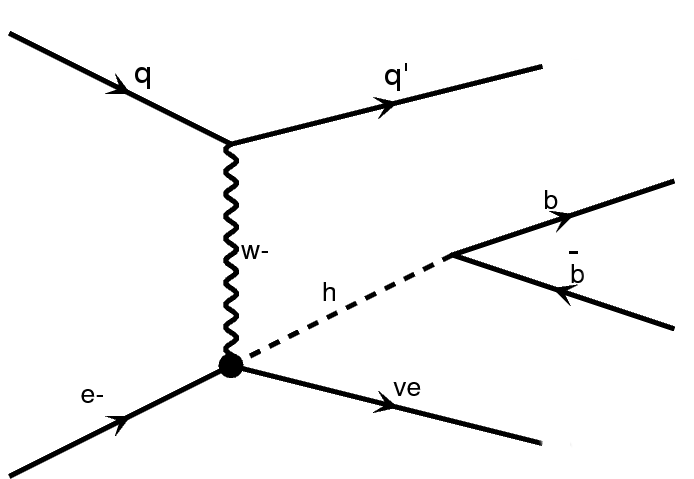}}
	\subfigure(c){\includegraphics[width=6.0cm]{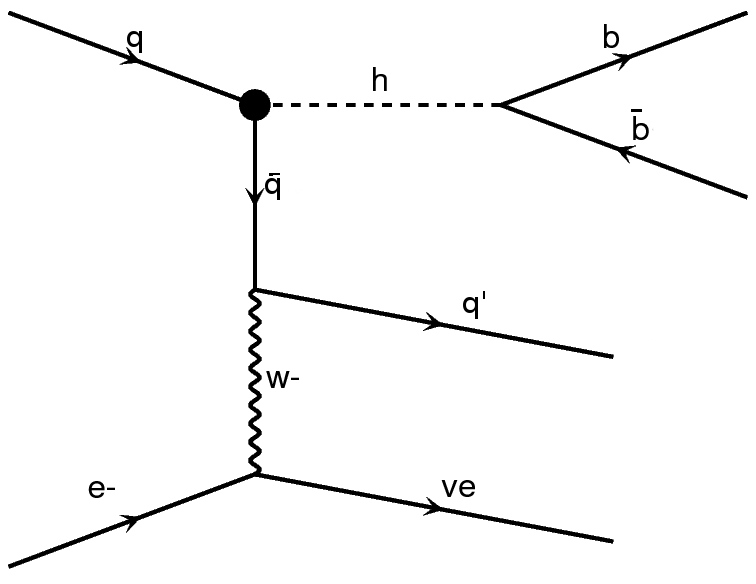}}
	\caption{ Representative Feynman diagrams at tree level for the $e^-p \to h j \nu_e$ process
	in electron-proton collisions in the presence of dimension six operators. }
	\label{fig:higgs_diagrameffective}
\end{figure}

The present work is dedicated to  consider the effects of ${\cal L}_{\rm eff}$ presented in Eq.~\ref{eq:effLag} on the
Higgs boson production through charged current process $e^{-}q\rightarrow Hq'\nu_{e}$.
The contributions originating from  other possible effective operators are neglected for simplicity.
The representative Feynman diagrams for $e^{-}q\rightarrow Hq'\nu_{e}$  are
displayed in Fig.~\ref{fig:higgs_diagrameffective}.
The vertices which receive contributions from the $\mathcal{L}_{\rm eff}$ are shown by
filled circles. It is remarkable that the SM tree level contribution does not have any dependency
on the momenta of the involved particles, while  considering $\mathcal{L}_{\rm eff}$
enters momentum-dependent interactions in the calculations.  This leads to changes in
the production cross sections as well as the shape of differential distributions.
In this paper,  differences in the shapes of distributions is used to constrain the involved Wilson coefficients in this process.

The $e^-p \to h j \nu_e$ process is sensitive to the following set of $\mathcal{L}_{\rm eff}$ parameters:
\begin{eqnarray}
	\bar{c}_{HW}, \tilde{c}_{HW}, \bar{c}_{W},\bar{c}_{H}, \bar{c}_{d},\bar{c}_{u},\bar{c}_{l}, \bar{c}_{Hud}, \bar{c^{\prime}}_{HL},  \bar{c^{\prime}}_{HQ},
	\bar{c}_{uW}, \bar{c}_{dW}, \bar{c}_{eW}\,,
\end{eqnarray}
The cross section of $e^-p \to h j \nu_e$ process is found to be almost insensitive to
parameters $\bar{c}_{l}$, $\bar{c}_{eW}$, $\bar{c}_{d},\bar{c}_{u},\bar{c}_{uW}, \bar{c}_{dW}$.
This is because of very small Yukawa couplings of light quarks and electron.
As a result, our analysis is restricted to the remaining seven parameters:

$$\bar{c}_{H}, \, \bar{c}_{Hud}, \, \bar{c}_{HW}, \, \bar{c^{\prime}}_{HL}, \, \bar{c^{\prime}}_{HQ}, \, \bar{c}_{W}, \, {\rm and} \, \tilde{c}_{HW} \,.$$

An interesting way to represent the effective Lagrangian from the experimental and phenomenological point of view
is the effective Lagrangian in the mass basis. In particular, it has been found to be an applicable approach in the
electroweak precision tests (EWPT) studies. The anomalous Higgs interactions in the mass basis
has been presented in Ref.~\cite{Alloul:2013naa}.
The relation between the mass basis couplings and the dimension-six coefficients which are involved in this analysis are given in Table~\ref{Tab1}.

\begin{table}[h!]
\caption{ Anomalous Higgs boson couplings in the mass basis and their relation with the dimension-six coefficients. }\label{Tab1}
	\begin{center}
		\begin{tabular}{c|c}
			\hline
			Mass basis          &  Gauge basis  		\\  \hline  \hline
			$g^{(1)}_{hww}$     & $\frac{2 g}{m_{W}} \bar c_{HW}$  \\ \hline
			$\tilde g_{hww}$    & $\frac{2 g}{m_{W}} \tilde c_{HW}$  \\ \hline
			$g^{(2)}_{hww}$     & $\frac{g}{m_{W}} \{\bar c_{W} + \bar c_{HW} \}$  \\ \hline
			$g_{hwud}^{(L)}$       &  $\frac{\sqrt{2} g}{v}  \bar c^\prime_{HQ} V^{\rm CKM}$  \\   \hline
			 $g_{hwud}^{(R)}$    & $\frac{\sqrt{2} g}{v}$  $\bar c_{Hud}$ \\   \hline
			  $g_{hw \nu e}$     &  $\frac{\sqrt{2} g}{v} \bar c^\prime_{HL}$   \\   \hline
		\end{tabular}
	\end{center}
\end{table}

There are already many studies to constrain the Wilson coefficients discussed above in different colliders using various channels
which can be found in
references~\cite{Hartmann:2016pil,Kuday:2017vsh,Kilian:2017nio,
Ellis:2017kfi,Fichet:2016iuo,Sigismondi:2012sp,Arbey:2016kqi,Amar:2014fpa,Banerjee:2015bla,
Craig:2015wwr,Corbett:2015ksa,Ellis:2014jta,Berthier:2015gja,Englert:2015hrx,Ellis:2014dva,
Khanpour:2017cfq,Khanpour:2017inb,Buckley:2015lku,Buckley:2015nca,Passarino:2016pzb,Giudice:2007fh,Ellis:2015sca,Bar-Shalom:2018rjs,Gu:2017ckc,Murphy:2017omb}.
Although the obtained limits on some of the coefficients in the previous studies are tight, 
we are going to examine possible improvements for these limits
in the future high-energy electron-proton colliders via a careful investigation of the
Higgs production mechanism in the framework of effective field theory. 
In the next section, the details of simulation for probing the effective Lagrangian
using  $e^-p \to h j \nu_e$ process in the future LHeC and FCC-he colliders will be  discussed.


\section{Signal and Backgrounds Production and Simulation}\label{sec:Simulation}

The chain we have used to perform the generation and simulation of the signal and background processes
are described is this section.
The full set of interactions generated by the dimension-six operators mentioned in the Higgs Effective
Lagrangian ${\cal L}_{\rm SILH}$ of Eq.~\eqref{eq:silh},  $\Delta {\cal L}_{F_1}$ in Eq.~\eqref{eq:silh2}
and $\Delta {\cal L}_{F_2}$ in Eq.~\eqref{eq:silh3} have been implemented
in {\tt FeynRules}~\cite{Christensen:2008py,Alloul:2013bka} and the model is imported to a
Universal {\tt FeynRules} Output (UFO) module~\cite{Degrande:2011ua,Alloul:2013naa}.
Then, the UFO model files have been inserted in the {\tt MadGraph5-aMC@NLO}~\cite{Alwall:2011uj,Alwall:2014hca}
Monte-Carlo (MC) event generator to calculate the cross sections and generate the signal events.
The {\tt CTEQ6L1} PDF set \cite{Pumplin:2002vw} is used to describe the proton structure functions.
The renormalization and factorization scales are set to be dynamical in {\tt MadGraph5-aMC@NLO}.

The next-to-leading order QCD correction to the signal process $e^-p \to h j \nu_e$
is found to be small~\cite{Jager:2010zm}. Therefore, in this work
the $k$-factor for the signal is assumed to be one.
The events of signal process $e^-p \to h j \nu_e$ are generated with {\tt MadGraph5-aMC@NLO},
then the Higgs boson decay into a $b\bar{b}$ pair is done with {\tt MadSpin}
module~\cite{Artoisenet:2012st,Frixione:2007zp}. {\tt Pythia 6}~\cite{Sjostrand:2003wg,Sjostrand:2007gs} package is
utilized to perform  fragmentation, hadronization, initial- and final-state parton showers.
Jets are clustered using {\tt FastJet3.2.0}~\cite{Cacciari:2011ma} with the $k_T$ algorithm~\cite{Soyez:2008pq}.

The electromagnetic and hadronic calorimeters resolutions are considered by the energy smearing
of $\frac{5\%}{\sqrt{E \, ({\rm GeV})}}$ (plus 1\% of constant term) and
 $\frac{60\%}{\sqrt{E \, ({\rm GeV})}}$, respectively~\cite{AbelleiraFernandez:2012cc}.
The b-tagging efficiency is assumed to be 60\% while mis-tag probabilities of $10\%$ and $1\%$  for $c$-quark jets and
light-quark jets are considered, respectively~\cite{AbelleiraFernandez:2012cc}.

The tracker of the LHeC detector is expected to cover pseudorapidity range up to 3.0~\cite{AbelleiraFernandez:2012cc}. Therefore,
 the $b$-tagging performance is valid up to $|\eta_{\rm b-jet}| < 3$.
For the light-jets, the calorimeter coverage is considered to be $|\eta_{\rm light-jet}| < 5$.

Based on the signal final state that consists of missing transverse energy,
a pair of $b\bar{b}$ from the Higgs boson decay and a forward  jet, the backgrounds include processes with three jets and large missing energy in the final state.
In particular,  the following processes have been taken into account:
$\, b b j^{\prime} \nu_{e}$, $\, b b b \nu_{e}$, $\, j^{\prime} j^{\prime} j^{\prime} \nu_{e}$, $\, t \nu_{e}$, $\, W j \nu_{e}$ and $Z j \nu_{e}$.
The hadronic decays of top quark, $W$-boson, and $Z$-boson are considered and $j^{\prime}$ refers to the light
flavors jets except the b-quark and $j$ denotes all light flavor quarks including b-quark. 
The background contributions from photo-production
processes which include the subprocesses $g\gamma \rightarrow b\bar{b}$ and $t\bar{t}$
are considered in the analysis.

In the next sections, we will present the analysis strategies for LHeC and FCC-he separately in more details.
As mentioned before, we consider the LHeC with the electron energies of 60 GeV and 140 GeV colliding the 7 TeV protons while 
for the FCC-he case, the 60 GeV electrons collide with 50 TeV protons.

%
\section{ LHeC sensitivity   } \label{sec:LHeC}

In this section, we present the analysis strategy and the results for the LHeC.
The strategy for choosing the basic cuts are similar to the one proposed in~\cite{AbelleiraFernandez:2012cc}.
Jets are reconstructed with a distance parameter for the jet reconstruction algorithm $R = 0.7$.
We require to have at least three jets with $p_T^{jets} > 20$ GeV from which
two are required to be $b$-tagged. A minimum cut of 20 GeV is imposed on the missing transverse energy
and the total transverse energy of all final state is required to be greater than 100 GeV.

\begin{figure}[htb]
 	\begin{center}
 		\vspace{0.5cm}
		\resizebox{0.49\textwidth}{!}{\includegraphics{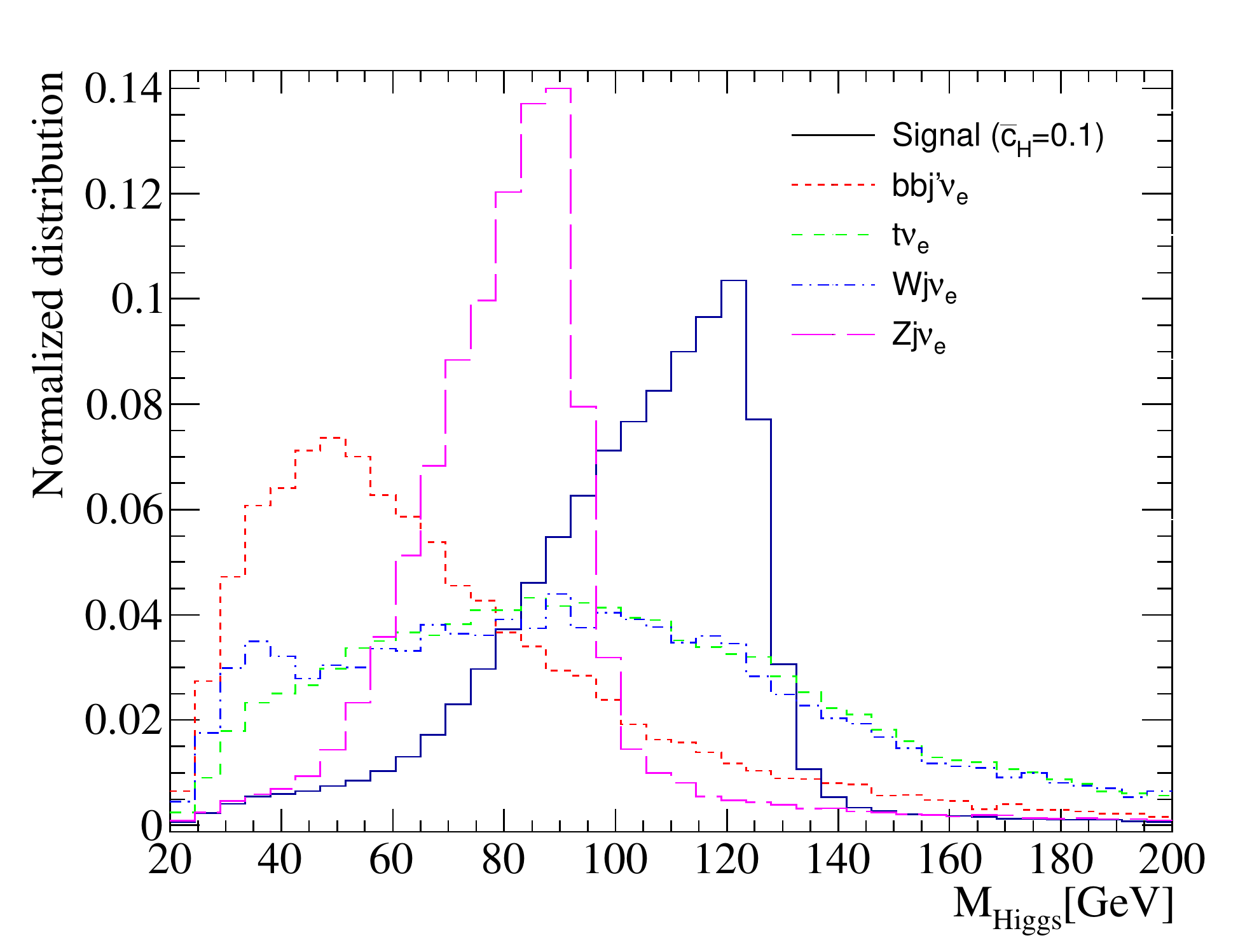}}
 		\resizebox{0.49\textwidth}{!}{\includegraphics{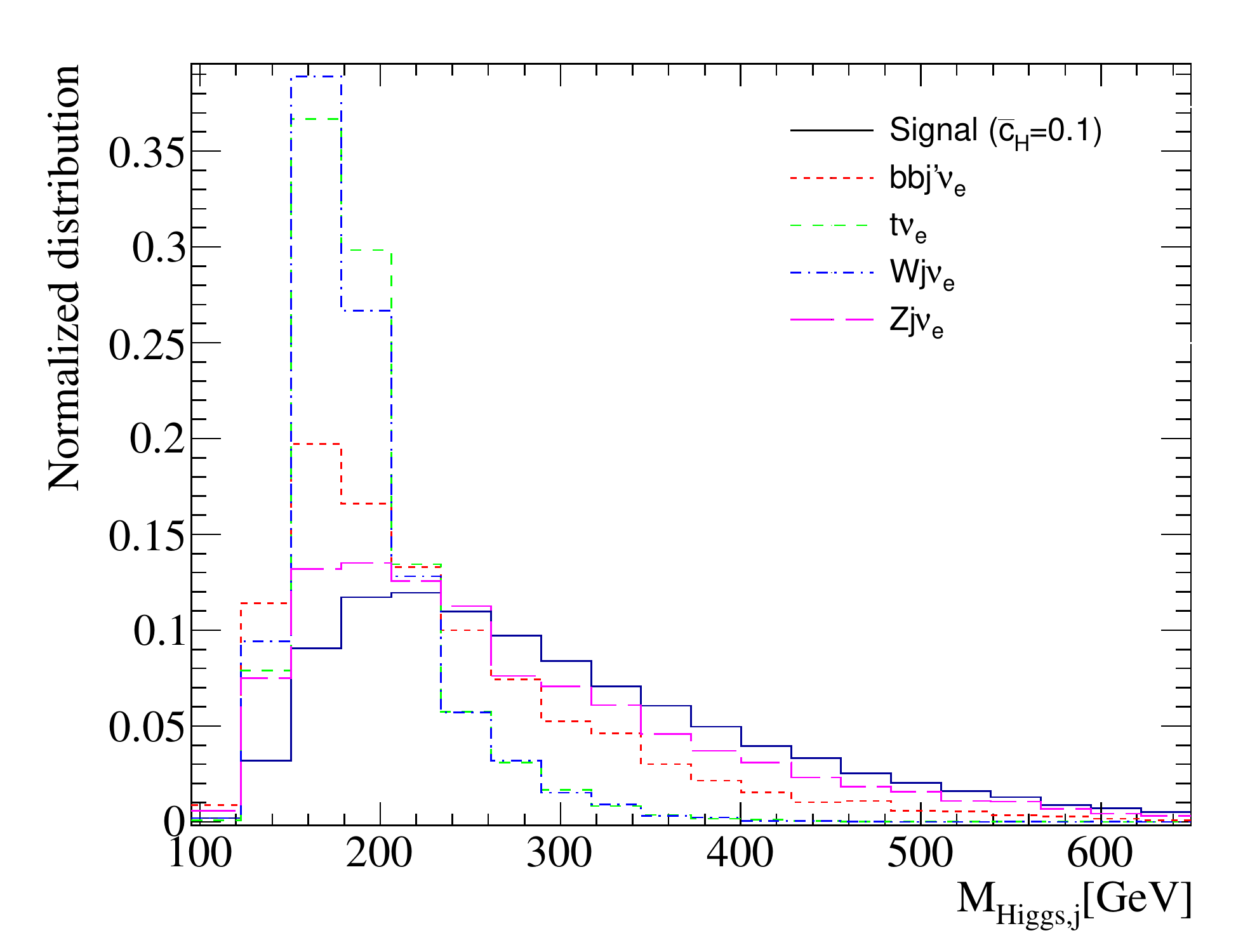}}
		\caption{ The Higgs boson invariant mass distribution (left) and invariant mass distributions of Higgs+jet (right) after the basic cuts. }
 		\label{fig:3jetsmass}
 	\end{center}
 \end{figure}

 The Higgs boson is reconstructed using the two $b$-tagged jets which give the closest mass to the nominal Higgs mass, i.e. 125 GeV. 
Among the light jets, the highest $p_T$ one is taken as the light flavor jet.
Figures~\ref{fig:3jetsmass} show the reconstructed Higgs boson mass (left) and the invariant mass distribution of the Higgs+jet ($M_{\rm Higgs,j}$) 
in the right panel for
signal with $\bar{c}_{H} = 0.1$ and the main backgrounds after the pre-selection cuts.

In the reconstructed distribution of the Higgs boson mass,
the mass peak is lower than the right Higgs boson mass because of 
 the energy carried by the neutrino from the $b$-quark decays.

The cross sections (in fb) after each cut for the signal, SM production of Higgs boson via $h j \nu_e$ process, and the main backgrounds
processes are presented in Table~\ref{Table:Cuts-CrossSection} for the LHeC operating with the electron energy of $E_e = 60 \, {\rm GeV}$. 
In addition to the basic kinematical cuts, a cut on the  invariant mass of the Higgs boosn $95 \leq M_{\rm Higgs} \leq 135$ GeV as well as 
the invariant mass of the Higgs-jet system
$260 < M_{\rm Higgs,j} <1000$ GeV
 are also applied for this analysis. 
 The impacts of these cuts are presented in Table~\ref{Table:Cuts-CrossSection}.
It clearly shows the selection criteria
are effective in enhancing  signal and suppression of the background contributions.
As one can see, cuts on Higgs boson invariant mass  and Higgs-jet system affect
all backgrounds and reduce their contributions significantly. The cross section of the photo-production
background after all cuts is found to be 0.18 fb. 

\begin{table*}[!]
\begin{tabular}{l|c|c|cccc}
LHeC collider  &  \multicolumn{1}{c|}{ Signal}  &  \multicolumn{1}{c|}{ Standard Model (SM)} & \multicolumn{4}{c}{Backgrounds }    \\   \hline
Cuts  &  $\bar{c}_{H}=0.1$ & $h j \nu_e$ &$bbj^{\prime} \nu_{e}$& $t \nu_{e}$ & $Wj\nu_{e}$ & $Zj\nu_{e}$  \\ 	\hline  \hline
Cross sections (in fb)	 &  $84.8$ &  $94.3$  & $639.5$ & $1287$ & $1885$ & $379.6$  \\  \hline
Acceptance cuts	 &  $18.10$ &  $20.12$  & $12.15$ & $96.76$ & $37.81$ & $16.33$  \\
$95 \leq M_{\rm Higgs} \leq 135 \, (\rm GeV)$  &  $9.69$ & $ 13.07$  & $1.28$ & $23.60$  & $10.08$  & $1.52$        \\
$ 260 < M_{\rm Higgs, j} <  1000 \, (\rm GeV)$  & $6.37$ & $7.05$   & $0.45$ & $1.77$  & $0.76$ & $0.72$     \\
\hline  \hline
\end{tabular}
\caption{  Cross section (in fb) for signal and background events after applied kinematic 
cuts used for this analysis at the LHeC with $E_e = 60 \, {\rm GeV}$. The details of the basic cuts applied, are presented in the text. }
\label{Table:Cuts-CrossSection}
\end{table*}

In this analysis, one might be  worried  for the  validity  of  the  effective field theory.
Several authors have discussed this issue in Refs.~\cite{Contino:2016jqw,Englert:2014cva,Farina:2016rws}. The Wilson coefficients of the dimension-six operators could be related to the 
new physics characteristic scale $M_{*}$ via 
\begin{eqnarray}
\bar{c} \sim \frac{g^{2}_{*}v^{2}}{M_{*}^2}\,,
\end{eqnarray}
where $g_{*}$ is the coupling constant of the heavy degrees of freedom with the SM particles.
Additional suppression factors appear in the case that an
operator is generated at loop level.  
An upper bound can be put on the new mass scale $M_{*}$ using the fact that 
the underlying  theory  is  strongly  coupled by setting $g_{*} = 4\pi$. Assuming $\bar{c} = \mathcal{O}(1)$, we find 
\begin{eqnarray}
M_{*} < \frac{4\pi v}{\sqrt{\bar{c}}} \sim 3.2~\text{TeV}\,.
\end{eqnarray} 
This upper bound is not violated in this analysis as we have $M_{\rm Higgs,j} < 1$ TeV.

%
\subsection{Sensitivity estimate } \label{sec:chi2-analysis}

\begin{figure}[htb]
	\begin{center}
		\vspace{0.5cm}
		\resizebox{0.70\textwidth}{!}{\includegraphics{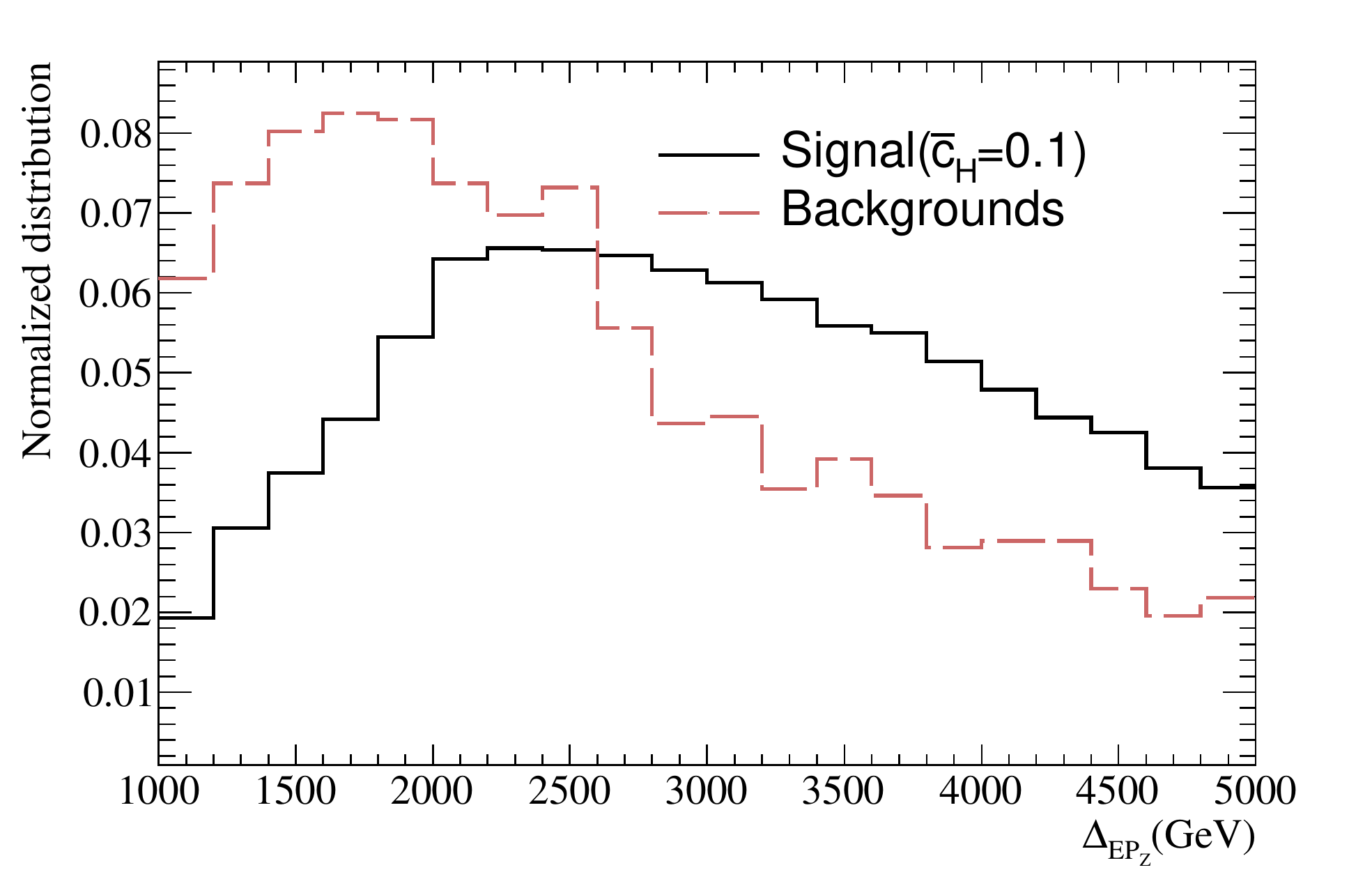}}  
		\caption{ Normalized distribution for the  $\Delta_{\rm Ep_{Z}}$ for signal and all background processes
		after applying all cuts presented in Table~\ref{Table:Cuts-CrossSection}.  }
		\label{fig:yjb}
	\end{center}
\end{figure}

This subsection is dedicated to estimate  the sensitivity of  $e^-p \to h j \nu_e$ process
to the Wilson coefficients. The sensitivity are obtained using a $\chi^{2}$ analysis over all bins of $\Delta_{E \rm p_{Z}}$ distribution.
The $\Delta_{E \rm p_{Z}}$ variable is defined as:
\begin{eqnarray}\label{DeltaEpZ}
\Delta_{E \rm p_{Z}} &=& (E_{{b-jet}_{1}}-p_{z,{b-jet}_{1}}) + (E_{{b-jet}_{2}}-p_{z,{b-jet}_{2}}) +   \nonumber  \\
&& (E_{{light-jet}_{1}}-p_{z,{light-jet}_{1}}) \,.
\end{eqnarray}
In Fig.~\ref{fig:yjb}, we show the the expected normalized distribution of $\Delta_{E \rm p_{Z}}$ for
the signal and the main sources of background processes after applying all cuts presented in Table~\ref{Table:Cuts-CrossSection}.
As it can be seen, the shapes $\Delta_{E \rm p_{Z}}$ signal with $\bar{c}_{H} =0.1$ is quite different
from the sum of all background processes. As a result, it is a useful distribution
to obtain the exclusion limits on the Wilson coefficients defined in Eq.\ref{eq:silh2} and Eq.\ref{eq:silh}.

To set upper limits at the $95\%$ CL, we use a $\chi^2$ criterion from the distribution of $\Delta_{E \rm p_{Z}}$ defined as:

\begin{eqnarray}\label{chi2}
\chi^{2}(\{ c_n \}) = \sum_{i = bins}^{N} \frac{(f_{i} (\{ c_n \}) - s^{\rm SM}_{i})^{2}}{\Delta_{i}^{2}} \,.
\end{eqnarray}

where $\{c_n\}$ denotes the Wilson coefficient $\{c_n = \bar{c}_{H},\, \bar{c}_{Hud}, \, \bar{c}_{HW},  \, \bar{c^{\prime}}_{HL},\,  \bar{c^{\prime}}_{HQ}, \, \bar{c}_{W}, \, \tilde{c}_{HW} \}$
considered in the present analysis, $s^{\rm SM}_{i}$ refers to the SM expectation in the $i$-th bin
of the $\Delta_{E \rm p_{Z}}$ distribution and $f_{i} (\{ c_n \})$ is number of signal events in the $i$-th bin.
In the $\chi^{2}(\{c_n\})$ definition, $\Delta_{i}$ is the statistical uncertainty.
We considered the most general formulation of $f_{i} (\{c_n\})$ as second-degree polynomials according to the following:

\begin{eqnarray} \label{si}
f_{i} (\{c_n\}) &=& s^{\rm SM}_{i} + \sum_{n=1}^{N} ( \alpha_{n} \bar{c}_{n} + \beta_{n}{\bar{c}_{n}}^{2} ) \,.
\end{eqnarray}

Considering only one coefficient in the fit, one can obtain exclusion limits for the individual constraints on
Wilson coefficient $ \{c_n= \bar{c}_{H}, \, \bar{c}_{Hud}, \, \bar{c}_{HW}, \, \bar{c^{\prime}}_{HL},
 \, \bar{c^{\prime}}_{HQ}, \, \bar{c}_{W}, \, \tilde{c}_{HW} \}$.
The corresponding results  for the integrated luminosities of 300 fb$^{-1}$, 3000 fb$^{-1}$ and 1 ab$^{-1}$
are presented in Table~\ref{tab:resultsLEC60-140}  and Table \ref{tab:resultsab} with the electron energies of 60 GeV and 140 GeV. 
As an example, LHeC would be able to constrain $\bar{c}_{H}$ by more than one order of magnitude with respect to
the LHC in high luminosity regime.

%
\section{  FCC-he sensitivity } \label{sec:FCC-he}

In this section, the sensitivity of the FCC-he to the related Wilson coefficients in the $h j \nu_{e}$ process
is studied. As we mentioned before, FCC-he employs the 50 TeV proton beam of a proposed circular proton-proton collider.

\begin{figure}[htb]
	\begin{center}
		\vspace{0.5cm}
		\resizebox{0.65\textwidth}{!}{\includegraphics{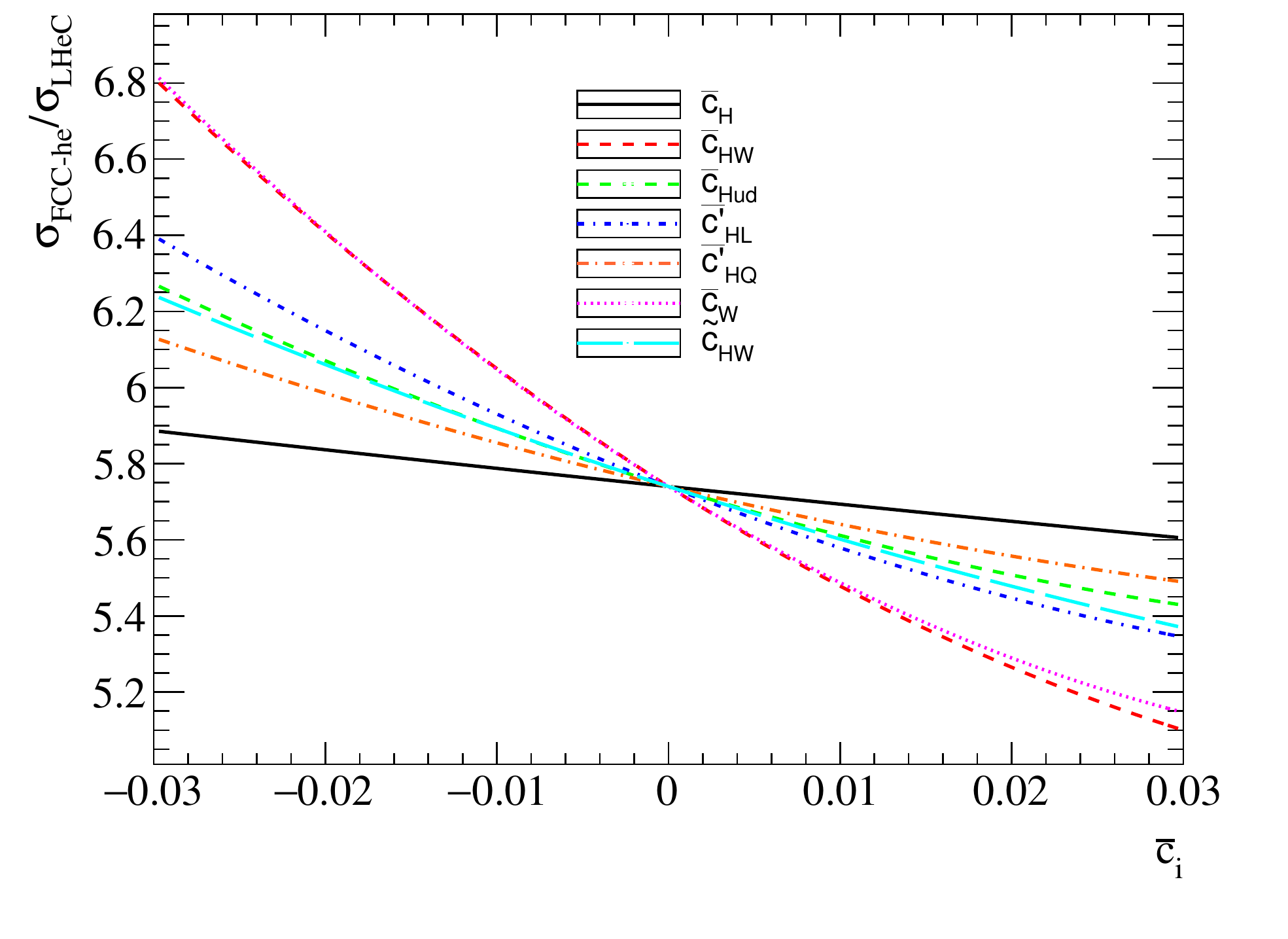}}  
		\caption{The ratio of signal cross section at the FCC-he ($E_{e} = 60$ GeV and $E_{p} = 50$ TeV)
		 to the LHeC ($E_{e} = 60$ GeV and $E_{p} = 7$ TeV) versus various Wilson coefficients.}
		\label{fig:ratio}
	\end{center}
\end{figure}

Similar to the LHeC case,  FCC-he is sensitive to $\bar{c}_{H}, \, \bar{c}_{Hud}, \, \bar{c}_{HW}, \, \bar{c^{\prime}}_{HL}, 
\, \bar{c^{\prime}}_{HQ}, \, \bar{c}_{W}, \, {\rm and} \, \tilde{c}_{HW}$ Wilson coefficients. The same  analysis strategy  as presented for the LHeC is followed for the FCC-he.
The Higgs boson decay into a $b \bar{b}$ pair is considered and  a $\chi^2$-fit is performed to estimate the sensitivities.
The ratio of the cross section of $e^-p \to h j \nu_e$  at the FCC-he to the LHeC in terms of the
Wilson coefficients is presented in Fig.\ref{fig:ratio}. As it can be seen, when the couplings are varying
in the range of -0.03 to 0.03, the cross section at the FCC-he increases by a factor of around 6 with respect to
the LHeC.

\begin{figure}[htb]
	\begin{center}
		\vspace{0.5cm}
		\resizebox{0.43\textwidth}{!}{\includegraphics{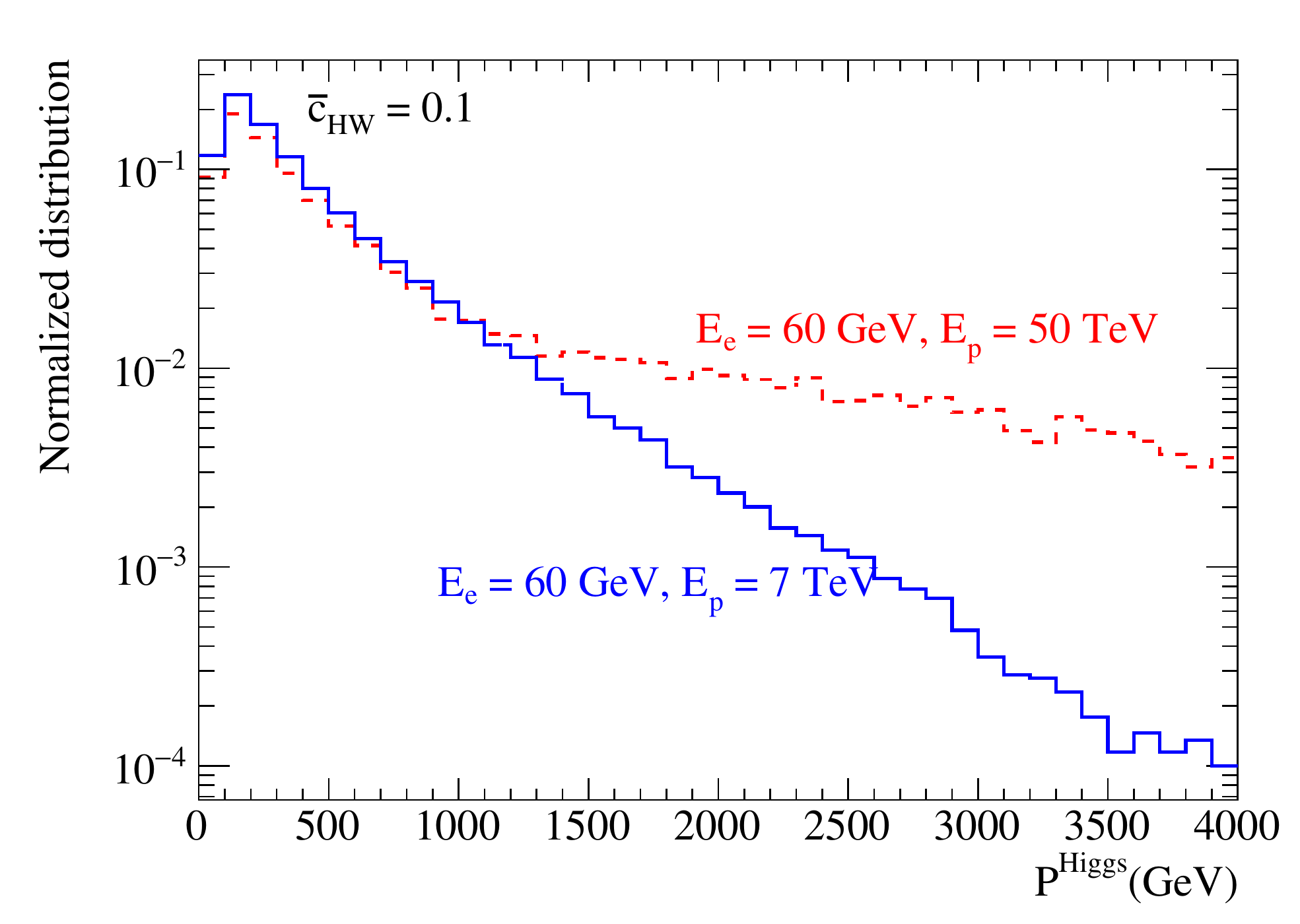}}  
	       \resizebox{0.4\textwidth}{!}{\includegraphics{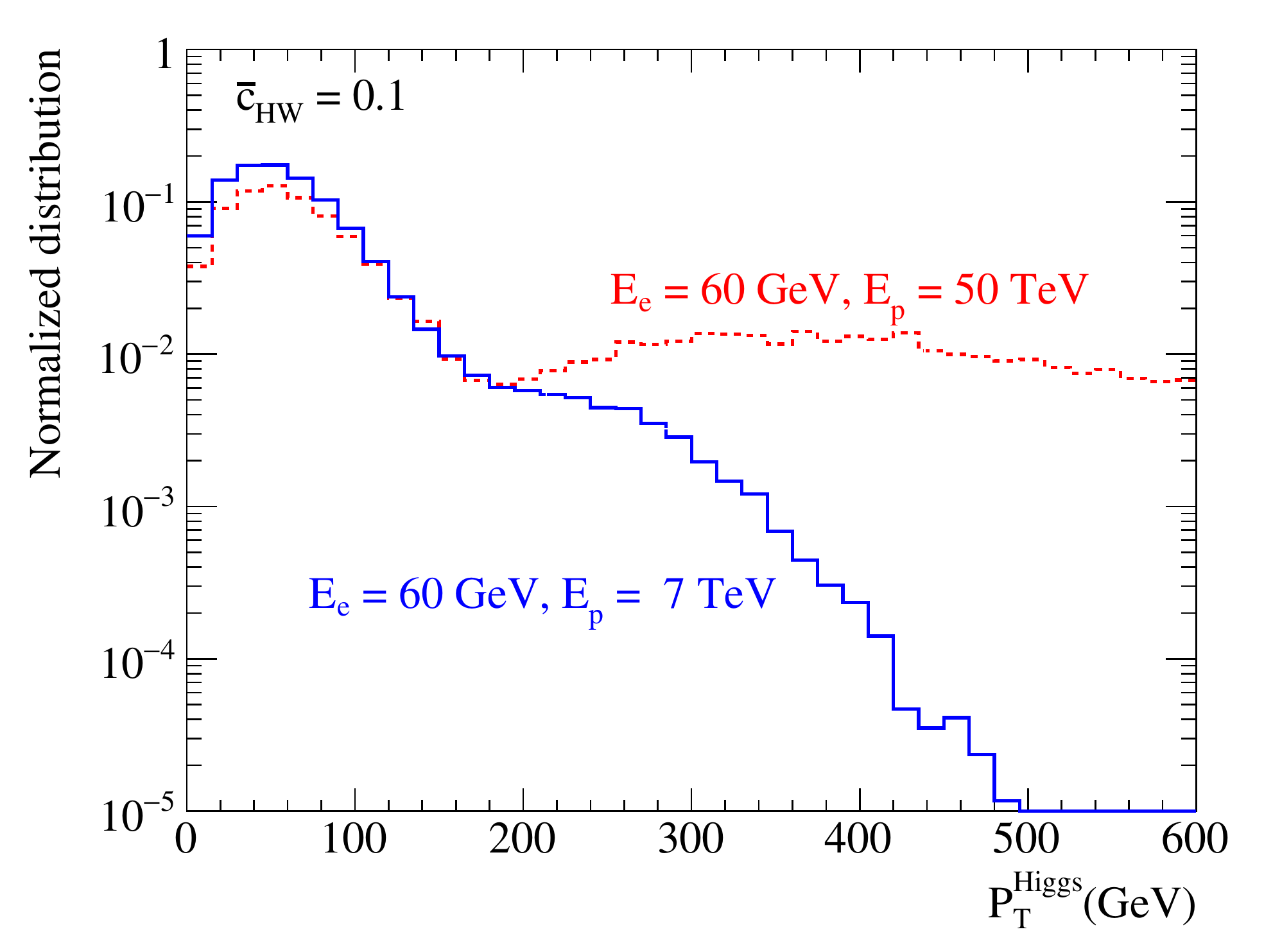}}  
		\caption{The distributions of the Higgs boson momentum (left) and the Higgs boson transverse momentum (right)  at the LHeC and FCC-he
		for the signal scenario of $\bar{c}_{HW} = 0.1$.}
		\label{fig:ppt}
	\end{center}
\end{figure}

At this step,  we mention one  of  the  interesting  characteristics  of  the  signal  events at the FCC-he,  which requires
to use a particular strategy for reconstruction of the Higgs boson.
At the FCC-he, Higgs bosons for some scenarios of signal are produced highly boosted
and  from the topological point of view they  decay  differently compared  to  the
Higgs bosons which are not boosted.  For instance, Higgs bosons produced in the signal scenario with non-zero value of $\bar{c}_{HW}$
are highly boosted because of the momentum dependent interaction. 
For the splitting of the Higgs boson into a pair of $b\bar{b}$, one can write:
\begin{eqnarray}\label{eq:deltar1}
m_{H}^{2} \simeq 2|\vec{p}_{b}||\vec{p}_{\bar{b}}|(1-\cos\theta_{b\bar{b}}),
\end{eqnarray}
where $\vec{p}_{b}(\vec{p}_{\bar{b}})$ is the momentum of the $b(\bar{b})$-quark and the bottom quark mass has been neglected. 
One can express the opening angle $\theta_{b\bar{b}}$ versus the parent mass Higgs boson and the momenta of the $b$- and 
$\bar{b}$-quark.  Using kinematic relations, the angular separation of a $b\bar{b}$ pair produced in a Higgs boson decay can be also written as:
\begin{equation}\label{eq:deltar}
\Delta R_{b \bar{b}} \simeq \frac{1}{\sqrt{x (1-x)}} \frac{m_{H}}{p_T} \,,
\end{equation}
where $p_{T}$ is the transverse momentum of the Higgs boson, $x$ and $1-x$ are the momentum fractions of the $b$ and $\bar{b}$ quarks.
Figure \ref{fig:ppt} shows the distributions of the Higgs boson momentum and transverse momentum for the 
signal scenario of $\bar{c}_{HW} = 0.1$ for the LHeC and FCC-he. As it can be seen, at the FCC-he 
Higgs bosons reside at large values of momentum and $p_{T}$.

\begin{figure}[!htbp]
	\begin{center}
		\includegraphics[width=7.0cm]{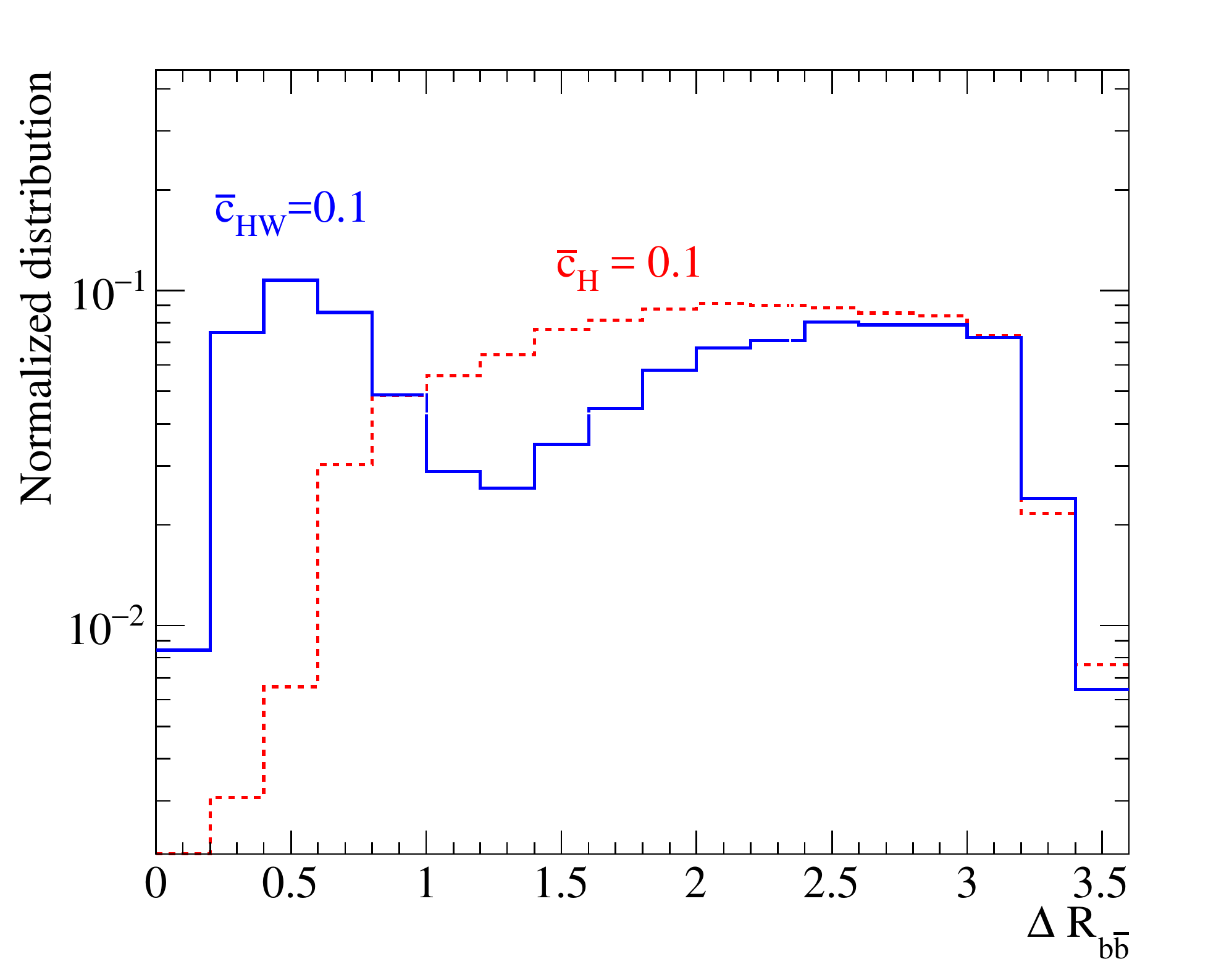}
		\hspace{5mm}
		\caption{ Normalized distribution of $\Delta R$ between two $b$-quarks coming from the decay of Higgs boson for two signal scenarios of 
		$\bar{c}_{H} = 0.1$ and $\bar{c}_{HW} = 0.1$ at the FCC-he. }
		\label{fig:drBB}
	\end{center}
\end{figure}

For the Higgs bosons with substantial momentum and $p_{T}$, from Eq.(\ref{eq:deltar1}) and Eq.(\ref{eq:deltar}) it is expected
that the angular separation of the Higgs boson decay products decreases.
Figure \ref{fig:drBB} shows the normalized distribution of $\Delta R$ between two $b$-quarks from the Higgs boson decay for the FCC-he.
We present the distributions for two  for two signal scenarios 
$\bar{c}_{H} = 0.1$ and $\bar{c}_{HW} = 0.1$.
The plot clearly confirms that for the signal scenario of $\bar{c}_{HW}$,
a considerable fraction of Higgs bosons are produced in the boosted regime while this
is not valid for the signal scenario of $\bar{c}_{H} $. As a result,  the high-$p_{T}$ Higgs bosons produce a
collimated jet with substructure.

Because of the
small angular separation between two $b$-jets from the Higgs decay and large boost,  the common jet reconstruction
 with a cone size of $\Delta R=0.4-0.5$ would not be usable
for most of the signal events with non-zero value of $\bar{c}_{HW}$. 
An alternative method of fat jet algorithm is applied~\cite{Butterworth:2008iy} for these boosted events.

To reconstruct the signal events with two boosted $b$-jets in the final state, first we reconstruct the fat jets by using the {\tt Cambridge/Aachen
(CA)} jet algorithm~\cite{Dokshitzer:1997in,Wobisch:1998wt} assuming a jet cone size of
$R=1.2$. Then, in order to identify the
boosted Higgs boson, the  methods described in the fat jet reconstruction algorithm~\cite{Butterworth:2008iy} is done as explained in the following.
In the beginning, a reconstructed fat jet $J$ is split into two sub-jets $J_{1}$,
$J_{2}$ with the masses $m_{J_{1}}>m_{J_{2}}$. Then, the method requires a significant mass drop
of $m_{J_{1}}<\mu_{MD} m_{J}$ with $\mu=0.667$. It should be mentioned that $\mu_{MD}$ is an arbitrary parameter
that shows the mass drop degree. Also, to avoid of including high $p_{T}$ light-jets,
two sub-jets are required to be symmetrically split. This requires the two sub-jets to satisfy:

\begin{equation}
\frac{min(p_{T,\, J_{1}}^{2},p_{T,\, J_{2}}^{2})}{m_{J}^{2}}\Delta R_{J_{1},\, J_{2}}^{2}<y_{cut}
\end{equation}

where $y_{cut}$ is a parameter of the algorithm which  determines the limit of asymmetry
between two sub-jets and $p_{T,\, J_{1}}^{2}$ and $p_{T,\, J_{2}}^{2}$ are the square
of the transverse momentum of each sub-jet.  Finally, if the above criteria are not satisfied,  the algorithm
takes $J=J_{1}$ and returns to the first step for performing decomposition.
The explained algorithm for boosted object reconstruction has been implemented in the {\tt FastJet} package~\cite{Cacciari:2011ma} by which our analysis is done.

In this analysis, first  jets of both the signal and backgrounds  are reconstructed with $k_{T}$ algorithm with a cone size of
$R=0.7$. Then if in an event, jets with $p_{T}>250$ GeV are found, the fat jet algorithm is applied otherwise, the event is considered as a normal event.
The exclusion limits for the Wilson coefficients of the dimension six operators at FCC-he are
obtained using a similar to what described in Section~\ref{sec:chi2-analysis}.

%
\section{ Results } \label{sec:results}

In this section the results are presented for the electron-proton collisions at the 
LHeC with the 60 GeV and 140 GeV electrons collide the 7 TeV protons, and at FCC-he with the60 GeV electrons collide the 50 TeV protons.
The limits are presented for the integrated luminosities of 300 fb$^{-1}$, 3000 fb$^{-1}$ and 1 ab$^{-1}$ 
in Tables~\ref{tab:resultsLEC60-140},  \ref{tab:results} and \ref{tab:resultsab}.

In Tables~\ref{tab:resultsLEC60-140} and \ref{tab:results},  we present the constraints on the Wilson coefficients that have been obtained at the LHC at 14 TeV with an integrated luminosity of 3000 fb$^{-1}$~\cite{Englert:2015hrx}.
As it can be seen from these tables, more sensitivity is achievable on the coefficient $\bar{c}_H$ at the electron-proton colliders with respect to the LHC.
Comparison between LHeC and FCC-he sensitivities shows that  more sensitivity to most of the Wilson coefficients can be obtained in FCC-he.
From these results, one can conclude that the LHeC and FCC-he are suitable platforms to complement the LHC results 
in search for dimension-six effective couplings in the Higgs boson sector.

\begin{table*}[htb]
	\begin{tabular}{l | C{2.5cm} | C{2.5cm} | C{2.5cm} | C{2.5cm}| C{2.5cm}}
		Wilson coefficients & LHeC-300 ($E_{e} = $ 140 GeV) & LHeC-3000 ($E_{e} = $ 140 GeV) & LHeC-300 ~  ($E_{e} = $ 60 GeV) & LHeC-3000 ($E_{e} = $ 60 GeV)  & LHC-3000~\cite{Englert:2015hrx}\\
		\hline
			\hline
		$\bar{c}_{H}[\times 100]$  &  $[\, -0.90, \, 0.95 \, ]$ &  $[\, -0.29, \, 0.29 \, ]$ &  $[\, -7.8, \, 8.8 \, ]$ &  $[\, -2.5, \, 2.6 \, ]$ & $[\, -4.40, \, 3.50 \, ]$ \\
		$\bar{c}_{Hud}[\times 100]$ &  $[\,-0.80, \, 0.80 \, ]$ &  $[\, -0.25, \, 0.25 \, ]$ & $[\, -6.26, \, 8.33 \, ]$   & $[\, -2.40, \, 2.86 \, ]$  & ------ \\
		$\bar{c}_{HW}[\times 100]$  &  $[\, -1.40, \,  1.70 \, ]$  &  $[\, -0.47, \, 0.50 \, ]$  & $[\, -2.3, \,  2.8 \, ]$  &  $[\, -0.79, \, 0.83 \, ]$  & $[\, -0.4, \, 0.4 \, ]$ \\
		$\bar{c^{\prime}}_{HL}[\times 100]$  &  $[\, -1.30, \, 1.40 \, ]$  &  $[\, -0.40, \,  0.40 \, ]$  & $[\, -2.6, \, 2.7 \, ]$   & $[\, -0.85, \, 0.82 \, ]$  & ------\\
		$\bar{c^{\prime}}_{HQ}[\times 100]$  &  $[\, -1.50, \,1.60 \, ]$ &  $[\, -0.50, \, 0.50 \, ]$ &$[\, -2.20, \,2.70 \, ]$   & $[\, -0.79, \, 0.76 \, ]$   & ------\\
		$\bar{c}_{W}[\times 100]$  &  $[\, -1.00, \, 1.00 \, ]$ &  $[\, -0.36, \, 0.37 \, ]$ & $[\, -1.20, \, 1.40 \, ]$  & $[\, -0.42, \, 0.44 \, ]$  &  $[\, -0.40, \, 0.40 \, ]$ \\
		$\tilde{c}_{HW}[\times 100]$  &  $[\, -0.70, \, 0.70 \, ]$  &  $[\, -0.20, \, 0.20 \, ]$  & $[\, -11.4, \, 9.2 \, ]$   &  $[\, -4.2, \, 3.6 \, ]$ & ------\\
		\hline
	\end{tabular}
	\caption{Predicted constraints at 95$\%$ C.L. on dimension-six Wilson coefficients for the LHeC with the electrons energy of  $E_e = 60 \, {\rm GeV}$ and $E_e = 140 \, {\rm GeV}$, and for integrated luminosities of $300~\text{fb}^{-1}$  and $3000~\text{fb}^{-1}$.
		\label{tab:resultsLEC60-140} }
\end{table*}

\begin{table*}[htb]
	\begin{tabular}{l | C{2.5cm} | C{2.5cm} | C{2.5cm} | C{2.5cm}| C{2.5cm}}
		Wilson coefficients & LHeC-300 ($E_{e} = $ 60 GeV) & LHeC-3000 ($E_{e} = $ 60 GeV) & FCC-he-300 ($E_{e} = $ 60 GeV)& FCC-he-3000 ( $E_{e} = $ 60 GeV)& LHC-3000~\cite{Englert:2015hrx}   \\
		\hline
		\hline
		$\bar{c}_{H}[\times 100]$ &  $[\, -7.8, \, 8.8 \, ]$ &  $[\, -2.5, \, 2.6 \, ]$ &  $[\, -8.70, \, 8.70 \, ]$ &  $[\, -2.75, \, 2.75 \, ]$ & $[\, -4.40, \, 3.50 \, ]$ \\
		$\bar{c}_{Hud}[\times 100]$ &$[\, -6.26, \, 8.33 \, ]$   & $[\, -2.40, \, 2.86 \, ]$  &  $[\, -4.00, \, 4.00 \, ]$&  $[\, -1.26, \, 1.26 \, ]$& ------ \\
		$\bar{c}_{HW}[\times 100]$  & $[\, -2.3, \,  2.8 \, ]$  &  $[\, -0.79, \, 0.83 \, ]$  & $[\, -1.00, \, 1.10 \, ]$&  $[\, -0.32, \, 0.35 \, ]$& $[\, -0.4, \, 0.4 \, ]$ \\
		$\bar{c^{\prime}}_{HL}[\times 100]$  &  $[\, -2.6, \, 2.7 \, ]$   & $[\, -0.85, \, 0.82 \, ]$    &  $[\, -4.50, \,  4.90 \, ]$  &  $[\, -1.42, \, 1.54 \, ]$ & ------\\
		$\bar{c^{\prime}}_{HQ}[\times 100]$  &  $[\, -2.20, \,2.70 \, ]$   & $[\, -0.79, \, 0.76 \, ]$  &  $[\, -5.70, \, 6.00 \, ]$ &  $[\, -1.80, \, 1.90 \, ]$ & ------\\
		$\bar{c}_{W}[\times 100]$  & $[\, -1.20, \, 1.40 \, ]$  & $[\, -0.42, \, 0.44 \, ]$  &  $[\, -1.20, \, 1.30 \, ]$ &  $[\, -0.38, \, 0.41 \, ]$ &  $[\, -0.40, \, 0.40 \, ]$\\
		$\tilde{c}_{HW}[\times 100]$  &  $[\, -11.4, \, 9.2 \, ]$   &  $[\, -4.2, \, 3.6 \, ]$  &  $[\, -4.70, \, 4.70 \, ]$  &  $[\, -1.49, \, 1.49 \, ]$ & ------\\
		\hline
	\end{tabular}
	\caption{ Predicted constraints at 95$\%$ C.L. on dimension-six Wilson coefficients for the LHeC and FCC-he colliders 
	and for integrated luminosity of $300~\text{fb}^{-1}$  and $3000~\text{fb}^{-1}$.
		\label{tab:results} }
\end{table*}

In order to study the effect arising from different energy of colliding electrons, we present the results for 
the LHeC with the electron energies of  $E_e = 60 \, {\rm GeV}$ and $E_e = 140 \, {\rm GeV}$.
From Table~\ref{tab:resultsLEC60-140}, it can be seen that going to higher energy of the electron-proton collisions, from 60 GeV to
140 GeV, would lead to improvements for the Wilson coefficients. 
For example, the constraints obtained from $E_e = 60 \, {\rm GeV}$ with the integrated luminosity of 3000 fb$^{-1}$ on
$\bar{c}_H$ is $-0.025 < \bar{c}_H < 0.026$ which is tightened to $-0.0029 < \bar{c}_H < 0.0029$ at a $E_e = 140 \, {\rm GeV}$ machine.

The results for LHeC and FCC-he at very high integrated luminosities are presented in Table~\ref{tab:resultsab}.
The bounds are given for maximum achievable integrated luminosities of 1 ab$^{-1}$ and 10 ab$^{-1}$ for the LHeC and FCC-he, respectively.
Based on this analysis for the FCC-he with $E_e = 60 \, {\rm GeV}$ for an integrated luminosity of 10 ab$^{-1}$, 
the sensitivity to the Wilson coefficients is much better than the other options analyzed in this study, 
and in some cases is  better than the ones expected to be achieved by the HL-LHC with an integrated luminosity of 3000 fb$^{-1}$.

\begin{table*}[htb]
	\begin{tabular}{l | C{5.0cm} | C{5.0cm} }
		Wilson coefficients & LHeC ($E_{e} = $ 60 GeV) 1 ab$^{-1}$ & FCC-he ( $E_{e} = $ 60 GeV) 10 ab$^{-1}$       \\
		\hline
		\hline
		$\bar{c}_{H}[\times 100]$  & $[\, -4.28, \, 4.60 \, ]$  & $[\, -1.5, \, 1.5 \, ]$   \\
		$\bar{c}_{Hud}[\times 100]$ & $[\, -3.88, \, 4.96 \, ]$  & $[\, -0.69, \, 0.69 \, ]$    \\
		$\bar{c}_{HW}[\times 100]$  & $[\, -0.89, \, 0.96 \, ]$   &   $[\, -0.17, \, 0.19 \, ]$   \\
		$\bar{c^{\prime}}_{HL}[\times 100]$  & $[\, -1.43, \, 1.58 \, ]$  & $[\, -0.78, \, 0.85 \, ]$   \\
		$\bar{c^{\prime}}_{HQ}[\times 100]$  & $[\, -1.29, \, 1.4 \, ]$  & $[\, -0.98, \, 1.01 \, ]$   \\
		$\bar{c}_{W}[\times 100]$  & $[\, -0.71, \, 0.76 \, ]$  & $[\, -0.21, \, 0.22 \, ]$   \\
		$\tilde{c}_{HW}[\times 100]$  &  $[\, -7.05, \, 6.03 \, ]$  & $[\, -0.81, \, 0.81 \, ]$   \\
		\hline
	\end{tabular}
	\caption{ Predicted constraints at 95$\%$ C.L. on Wilson coefficients for the LHeC  and FCC-he with $E_e = 60 \, {\rm GeV}$. 
	 The limits presented for the integrated luminosities of $1~\text{ab}^{-1}$ for LHeC and $10~\text{ab}^{-1}$ for FCC-he, respectively.
	\label{tab:resultsab} }
\end{table*}

%
\section{Summary and conclusions} \label{sec:Discussion}
%

The effects of physics beyond SM may appear in the Higgs sector  which requires to measure the Higgs boson couplings with the SM particles precisely.
Any deviation of the Higgs boson interactions with respect to the predictions of the SM would be a hint to new physics.
The Large Hadron Electron Collider (LHeC) with a rich physics program would be able to provide
a lot of information on physics beyond the SM as well as providing precise measurements of the SM.
In electron-proton collisions, there are two clean production mechanisms for the Higgs boson
either in neutral current interactions or in charged current interactions.
In this paper we present an analysis to constrain new physics in the Higgs boson sector by adopting an effective Lagrangian approach.
The analysis is based on the Higgs boson production in charged current interactions (via the WWH coupling)
i.e.,  $e^- p \to h j \nu_{e}$ process for the electrons with the energies of $E_e = 60 \, {\rm GeV}$ and $E_e = 140 \, {\rm GeV}$
colliding the 7 TeV protons.
We also perform the same analysis for the Future Circular Hadron Electron Collider (FCC-he) in
which 60 GeV electrons collide very high energy protons with the energy of 50 TeV.
For the FCC-he case,  to efficiently  reconstruct the Higgs boson and to achieve a reasonable background
rejection, jet substructure techniques are used in order to capture the signal events which are boosted
objects.

To obtain the sensitivity to the involved Wilson coefficients of dimension-six operators, an analysis on the
kinematic distribution of $\Delta_{E \rm p_{Z}}$ (defined in Eq.~\eqref{DeltaEpZ}) is performed.
The Higgs boson production via charged current interaction, $e^- p \to h j \nu_{e}$,  is in particular sensitive to a variety of Wilson coefficients namely
$\{ c_i = \bar{c}_{H},\, \bar{c}_{Hud}, \, \bar{c}_{HW},  \, \bar{c^{\prime}}_{HL},\,  \bar{c^{\prime}}_{HQ}, \, \bar{c}_{W}, \, \tilde{c}_{HW} \}$.
The extracted bounds for both LHeC and FCC-he which are presented in Table~\ref{tab:results} showing a great sensitivity and in some cases
improvements are expected with respect to the potential constraints for the LHC~\cite{Englert:2015hrx,Khanpour:2017inb}.
We also show that the FCC-he collider with $E_e = 60 \, {\rm GeV}$ and 
with an integrated luminosity of $\cal L =$ 10 ab$^{-1}$ or even with 3 ab$^{-1}$ 
would be able to probe the Wilson coefficients of dimension-six operators of the Higgs boson 
(especially $\bar{c}_{H}$, $\bar{c}_{HW}$ and $\bar{c}_{W}$ couplings) beyond the HL-LHC.

%
\section*{Acknowledgments}
%

Authors thank School of Particles and Accelerators, 
Institute for Research in Fundamental Sciences (IPM) for financial support of this project. Hamzeh Khanpour also is grateful to
the University of Science and Technology of Mazandaran for financial support provided for this research. M. Mohammadi Najafabadi
is thankful to the Iran National Science Foundation (INSF).

%

\begin{thebibliography}{}
%


\bibitem{Aad:2012tfa} 
G.~Aad {\it et al.} [ATLAS Collaboration],
``Observation of a new particle in the search for the Standard Model Higgs boson with the ATLAS detector at the LHC,''
Phys.\ Lett.\ B {\bf 716}, 1 (2012)
doi:10.1016/j.physletb.2012.08.020
[arXiv:1207.7214 [hep-ex]].





\bibitem{Chatrchyan:2012xdj} 
S.~Chatrchyan {\it et al.} [CMS Collaboration],
``Observation of a new boson at a mass of 125 GeV with the CMS experiment at the LHC,''
Phys.\ Lett.\ B {\bf 716}, 30 (2012)
doi:10.1016/j.physletb.2012.08.021
[arXiv:1207.7235 [hep-ex]].



\bibitem{Buchmuller:1985jz} 
W.~Buchmuller and D.~Wyler,
``Effective Lagrangian Analysis of New Interactions and Flavor Conservation,''
Nucl.\ Phys.\ B {\bf 268}, 621 (1986).
doi:10.1016/0550-3213(86)90262-2




\bibitem{Grzadkowski:2010es} 
B.~Grzadkowski, M.~Iskrzynski, M.~Misiak and J.~Rosiek,
``Dimension-Six Terms in the Standard Model Lagrangian,''
JHEP {\bf 1010}, 085 (2010)
doi:10.1007/JHEP10(2010)085
[arXiv:1008.4884 [hep-ph]].





\bibitem{AguilarSaavedra:2008zc} 
J.~A.~Aguilar-Saavedra,
``A Minimal set of top anomalous couplings,''
Nucl.\ Phys.\ B {\bf 812}, 181 (2009)
doi:10.1016/j.nuclphysb.2008.12.012
[arXiv:0811.3842 [hep-ph]].





\bibitem{Arzt:1994gp} 
C.~Arzt, M.~B.~Einhorn and J.~Wudka,
``Patterns of deviation from the standard model,''
Nucl.\ Phys.\ B {\bf 433}, 41 (1995)
doi:10.1016/0550-3213(94)00336-D
[hep-ph/9405214].






\bibitem{Hartmann:2016pil} 
C.~Hartmann, W.~Shepherd and M.~Trott,
``The $Z$ decay width in the SMEFT: $y_t$ and $\lambda$ corrections at one loop,''
JHEP {\bf 1703}, 060 (2017)
doi:10.1007/JHEP03(2017)060
[arXiv:1611.09879 [hep-ph]].





\bibitem{Kuday:2017vsh} 
S.~Kuday, H.~Saygin, I.~Hos and F.~Cetin,
``Limits on Neutral Di-Boson and Di-Higgs Interactions for FCC-he Collider,''
arXiv:1702.00185 [hep-ph].





\bibitem{Kilian:2017nio} 
W.~Kilian, S.~Sun, Q.~S.~Yan, X.~Zhao and Z.~Zhao,
``New Physics in multi-Higgs boson final states,''
JHEP {\bf 1706}, 145 (2017)
doi:10.1007/JHEP06(2017)145
[arXiv:1702.03554 [hep-ph]].





\bibitem{Ellis:2017kfi} 
J.~Ellis, P.~Roloff, V.~Sanz and T.~,
``Dimension-6 Operator Analysis of the CLIC Sensitivity to New Physics,''
JHEP {\bf 1705}, 096 (2017)
doi:10.1007/JHEP05(2017)096
[arXiv:1701.04804 [hep-ph]].




\bibitem{Fichet:2016iuo} 
S.~Fichet, A.~Tonero and P.~Rebello Teles,
``Sharpening the shape analysis for higher-dimensional operator searches,''
Phys.\ Rev.\ D {\bf 96}, no. 3, 036003 (2017)
doi:10.1103/PhysRevD.96.036003
[arXiv:1611.01165 [hep-ph]].





\bibitem{Sigismondi:2012sp} 
C.~Sigismondi,
``Measuring the position of the center of the Sun at the Clementine Gnomon of Santa Maria degli Angeli in Rome,''
J.\ Occult.\ Astron.\  {\bf 1N5}, 20 (2012)
[arXiv:1201.0510 [astro-ph.IM]].





\bibitem{Arbey:2016kqi} 
A.~Arbey, S.~Fichet, F.~Mahmoudi and G.~Moreau,
``The correlation matrix of Higgs rates at the LHC,''
JHEP {\bf 1611}, 097 (2016)
doi:10.1007/JHEP11(2016)097
[arXiv:1606.00455 [hep-ph]].




\bibitem{Amar:2014fpa} 
G.~Amar, S.~Banerjee, S.~von Buddenbrock, A.~S.~Cornell, T.~Mandal, B.~Mellado and B.~Mukhopadhyaya,
``Exploration of the tensor structure of the Higgs boson coupling to weak bosons in e$^{+}$ e$^{-}$ collisions,''
JHEP {\bf 1502}, 128 (2015)
doi:10.1007/JHEP02(2015)128
[arXiv:1405.3957 [hep-ph]].




\bibitem{Banerjee:2015bla} 
S.~Banerjee, T.~Mandal, B.~Mellado and B.~Mukhopadhyaya,
``Cornering dimension-6 $HVV$ interactions at high luminosity LHC: the role of event ratios,''
JHEP {\bf 1509}, 057 (2015)
doi:10.1007/JHEP09(2015)057
[arXiv:1505.00226 [hep-ph]].




\bibitem{Craig:2015wwr} 
N.~Craig, J.~Gu, Z.~Liu and K.~Wang,
``Beyond Higgs Couplings: Probing the Higgs with Angular Observables at Future e$^{+}$ e$^{-}$ Colliders,''
JHEP {\bf 1603}, 050 (2016)
doi:10.1007/JHEP03(2016)050
[arXiv:1512.06877 [hep-ph]].




\bibitem{Corbett:2015ksa} 
T.~Corbett, O.~J.~P.~Eboli, D.~Goncalves, J.~Gonzalez-Fraile, T.~Plehn and M.~Rauch,
``The Higgs Legacy of the LHC Run I,''
JHEP {\bf 1508}, 156 (2015)
doi:10.1007/JHEP08(2015)156
[arXiv:1505.05516 [hep-ph]].





\bibitem{Ellis:2014jta} 
J.~Ellis, V.~Sanz and T.~You,
``The Effective Standard Model after LHC Run I,''
JHEP {\bf 1503}, 157 (2015)
doi:10.1007/JHEP03(2015)157
[arXiv:1410.7703 [hep-ph]].




\bibitem{Berthier:2015gja} 
L.~Berthier and M.~Trott,
``Consistent constraints on the Standard Model Effective Field Theory,''
JHEP {\bf 1602}, 069 (2016)
doi:10.1007/JHEP02(2016)069
[arXiv:1508.05060 [hep-ph]].




\bibitem{Englert:2015hrx} 
C.~Englert, R.~Kogler, H.~Schulz and M.~Spannowsky,
``Higgs coupling measurements at the LHC,''
Eur.\ Phys.\ J.\ C {\bf 76}, no. 7, 393 (2016)
doi:10.1140/epjc/s10052-016-4227-1
[arXiv:1511.05170 [hep-ph]].




\bibitem{Ellis:2014dva} 
J.~Ellis, V.~Sanz and T.~You,
``Complete Higgs Sector Constraints on Dimension-6 Operators,''
JHEP {\bf 1407}, 036 (2014)
doi:10.1007/JHEP07(2014)036
[arXiv:1404.3667 [hep-ph]].




\bibitem{Khanpour:2017cfq} 
H.~Khanpour and M.~Mohammadi Najafabadi,
``Constraining Higgs boson effective couplings at electron-positron colliders,''
Phys.\ Rev.\ D {\bf 95}, no. 5, 055026 (2017)
doi:10.1103/PhysRevD.95.055026
[arXiv:1702.00951 [hep-ph]].





\bibitem{Khanpour:2017inb} 
H.~Khanpour, S.~Khatibi and M.~Mohammadi Najafabadi,
``Probing Higgs boson couplings in H+$\gamma$ production at the LHC,''
Phys.\ Lett.\ B {\bf 773}, 462 (2017)
doi:10.1016/j.physletb.2017.09.005
[arXiv:1702.05753 [hep-ph]].





\bibitem{Buckley:2015lku} 
A.~Buckley, C.~Englert, J.~Ferrando, D.~J.~Miller, L.~Moore, M.~Russell and C.~D.~White,
``Constraining top quark effective theory in the LHC Run II era,''
JHEP {\bf 1604}, 015 (2016)
doi:10.1007/JHEP04(2016)015
[arXiv:1512.03360 [hep-ph]].



\bibitem{Buckley:2015nca} 
A.~Buckley, C.~Englert, J.~Ferrando, D.~J.~Miller, L.~Moore, M.~Russell and C.~D.~White,
``Global fit of top quark effective theory to data,''
Phys.\ Rev.\ D {\bf 92}, no. 9, 091501 (2015)
doi:10.1103/PhysRevD.92.091501
[arXiv:1506.08845 [hep-ph]].




\bibitem{Denizli:2017pyu} 
H.~Denizli and A.~Senol,
``Constraints on Higgs effective couplings in $H\nu \bar{\nu}$ production of CLIC at 380 GeV,''
Adv.\ High Energy Phys.\  {\bf 2018}, 1627051 (2018)
doi:10.1155/2018/1627051
[arXiv:1707.03890 [hep-ph]].





\bibitem{Barklow:2017suo} 
T.~Barklow, K.~Fujii, S.~Jung, R.~Karl, J.~List, T.~Ogawa, M.~E.~Peskin and J.~Tian,
``Improved Formalism for Precision Higgs Coupling Fits,''
arXiv:1708.08912 [hep-ph].




\bibitem{Murphy:2017omb} 
C.~W.~Murphy,
``Statistical approach to Higgs boson couplings in the standard model effective field theory,''
Phys.\ Rev.\ D {\bf 97}, no. 1, 015007 (2018)
doi:10.1103/PhysRevD.97.015007
[arXiv:1710.02008 [hep-ph]].




\bibitem{Jana:2017hqg} 
S.~Jana and S.~Nandi,
``New Physics Scale from Higgs Observables with Effective Dimension-6 Operators,''
arXiv:1710.00619 [hep-ph].


\bibitem{Dedes:2017zog} 
A.~Dedes, W.~Materkowska, M.~Paraskevas, J.~Rosiek and K.~Suxho,
``Feynman rules for the Standard Model Effective Field Theory in R$_{?}$ -gauges,''
JHEP {\bf 1706}, 143 (2017)
doi:10.1007/JHEP06(2017)143
[arXiv:1704.03888 [hep-ph]].



\bibitem{Dedes:2018seb} 
A.~Dedes, M.~Paraskevas, J.~Rosiek, K.~Suxho and L.~Trifyllis,
``The decay $h\to \gamma\gamma$ in the Standard-Model Effective Field Theory,''
arXiv:1805.00302 [hep-ph].







\bibitem{Contino:2013kra} 
R.~Contino, M.~Ghezzi, C.~Grojean, M.~Muhlleitner and M.~Spira,
``Effective Lagrangian for a light Higgs-like scalar,''
JHEP {\bf 1307}, 035 (2013)
doi:10.1007/JHEP07(2013)035
[arXiv:1303.3876 [hep-ph]].






\bibitem{Alloul:2013naa} 
A.~Alloul, B.~Fuks and V.~Sanz,
``Phenomenology of the Higgs Effective Lagrangian via FEYNRULES,''
JHEP {\bf 1404}, 110 (2014)
doi:10.1007/JHEP04(2014)110
[arXiv:1310.5150 [hep-ph]].





\bibitem{Artoisenet:2013puc} 
P.~Artoisenet {\it et al.},
``A framework for Higgs characterisation,''
JHEP {\bf 1311}, 043 (2013)
doi:10.1007/JHEP11(2013)043
[arXiv:1306.6464 [hep-ph]].





\bibitem{Bruening:2013bga} 
O.~Bruening and M.~Klein,
``The Large Hadron Electron Collider,''
Mod.\ Phys.\ Lett.\ A {\bf 28}, no. 16, 1330011 (2013)
doi:10.1142/S0217732313300115
[arXiv:1305.2090 [physics.acc-ph]].





\bibitem{AbelleiraFernandez:2012cc} 
J.~L.~Abelleira Fernandez {\it et al.} [LHeC Study Group],
``A Large Hadron Electron Collider at CERN: Report on the Physics and Design Concepts for Machine and Detector,''
J.\ Phys.\ G {\bf 39}, 075001 (2012)
doi:10.1088/0954-3899/39/7/075001
[arXiv:1206.2913 [physics.acc-ph]].



\bibitem{AbelleiraFernandez:2012ty} 
J.~L.~Abelleira Fernandez {\it et al.} [LHeC Study Group],
``On the Relation of the LHeC and the LHC,''
arXiv:1211.5102 [hep-ex].




\bibitem{AbelleiraFernandez:2012ni} 
J.~L.~Abelleira Fernandez {\it et al.},
``A Large Hadron Electron Collider at CERN,''
arXiv:1211.4831 [hep-ex].




\bibitem{Kumar:2015kca} 
M.~Kumar, X.~Ruan, R.~Islam, A.~S.~Cornell, M.~Klein, U.~Klein and B.~Mellado,
``Probing anomalous couplings using di-Higgs production in electron-proton collisions,''
Phys.\ Lett.\ B {\bf 764}, 247 (2017)
doi:10.1016/j.physletb.2016.11.039
[arXiv:1509.04016 [hep-ph]].






\bibitem{Pomarol:2013zra} 
A.~Pomarol and F.~Riva,
``Towards the Ultimate SM Fit to Close in on Higgs Physics,''
JHEP {\bf 1401}, 151 (2014)
doi:10.1007/JHEP01(2014)151
[arXiv:1308.2803 [hep-ph]].




\bibitem{Ellis:1975ap} 
J.~R.~Ellis, M.~K.~Gaillard and D.~V.~Nanopoulos,
``A Phenomenological Profile of the Higgs Boson,''
Nucl.\ Phys.\ B {\bf 106}, 292 (1976).
doi:10.1016/0550-3213(76)90382-5




\bibitem{LoSecco:1976ii} 
J.~M.~LoSecco,
``Higgs Boson Production in Neutrino Scattering,''
Phys.\ Rev.\ D {\bf 14}, 1352 (1976).
doi:10.1103/PhysRevD.14.1352




\bibitem{Hioki:1983yz} 
Z.~Hioki, S.~Midorikawa and H.~Nishiura,
``Higgs Boson Production in High-energy Lepton - Nucleon Scattering,''
Prog.\ Theor.\ Phys.\  {\bf 69}, 1484 (1983).
doi:10.1143/PTP.69.1484



\bibitem{Blumlein:1992eh} 
J.~Blumlein, G.~J.~van Oldenborgh and R.~Ruckl,
``QCD and QED corrections to Higgs boson production in charged current e p scattering,''
Nucl.\ Phys.\ B {\bf 395}, 35 (1993)
doi:10.1016/0550-3213(93)90207-6
[hep-ph/9209219].




\bibitem{Han:2009pe} 
T.~Han and B.~Mellado,
``Higgs Boson Searches and the H b anti-b Coupling at the LHeC,''
Phys.\ Rev.\ D {\bf 82}, 016009 (2010)
doi:10.1103/PhysRevD.82.016009
[arXiv:0909.2460 [hep-ph]].





\bibitem{Biswal:2012mp} 
S.~S.~Biswal, R.~M.~Godbole, B.~Mellado and S.~Raychaudhuri,
``Azimuthal Angle Probe of Anomalous $HWW$ Couplings at a High Energy $ep$ Collider,''
Phys.\ Rev.\ Lett.\  {\bf 109}, 261801 (2012)
doi:10.1103/PhysRevLett.109.261801
[arXiv:1203.6285 [hep-ph]].



\bibitem{Sun:2016kek} 
H.~Sun and X.~Wang,
``Searches for the Anomalous FCNC Top-Higgs Couplings at the LHeC,''
arXiv:1602.04670 [hep-ph].





\bibitem{Wang:2017pdg} 
X.~Wang, H.~Sun and X.~Luo,
``Searches for the Anomalous FCNC Top-Higgs Couplings with Polarized Electron Beam at the LHeC,''
Adv.\ High Energy Phys.\  {\bf 2017}, 4693213 (2017)
doi:10.1155/2017/4693213
[arXiv:1703.02691 [hep-ph]].






\bibitem{Senol:2012fc} 
A.~Senol,
``Anomalous Higgs Couplings at the LHeC,''
Nucl.\ Phys.\ B {\bf 873}, 293 (2013)
doi:10.1016/j.nuclphysb.2013.04.016
[arXiv:1212.6869 [hep-ph]].





\bibitem{Passarino:2016pzb} 
G.~Passarino and M.~Trott,
``The Standard Model Effective Field Theory and Next to Leading Order,''
arXiv:1610.08356 [hep-ph].





\bibitem{Giudice:2007fh} 
G.~F.~Giudice, C.~Grojean, A.~Pomarol and R.~Rattazzi,
``The Strongly-Interacting Light Higgs,''
JHEP {\bf 0706}, 045 (2007)
doi:10.1088/1126-6708/2007/06/045
[hep-ph/0703164].



\bibitem{Ellis:2015sca} 
J.~Ellis and T.~You,
``Sensitivities of Prospective Future e+e- Colliders to Decoupled New Physics,''
JHEP {\bf 1603}, 089 (2016)
doi:10.1007/JHEP03(2016)089
[arXiv:1510.04561 [hep-ph]].

\bibitem{Bar-Shalom:2018rjs} 
  S.~Bar-Shalom and A.~Soni,
  ``A universally enhanced light-quarks Yukawa couplings paradigm,''
  arXiv:1804.02400 [hep-ph].

\bibitem{Gu:2017ckc} 
  J.~Gu, H.~Li, Z.~Liu, S.~Su and W.~Su,
  ``Learning from Higgs Physics at Future Higgs Factories,''
  JHEP {\bf 1712}, 153 (2017)
  doi:10.1007/JHEP12(2017)153
  [arXiv:1709.06103 [hep-ph]].





\bibitem{Christensen:2008py} 
N.~D.~Christensen and C.~Duhr,
``FeynRules - Feynman rules made easy,''
Comput.\ Phys.\ Commun.\  {\bf 180}, 1614 (2009)
doi:10.1016/j.cpc.2009.02.018
[arXiv:0806.4194 [hep-ph]].




\bibitem{Alloul:2013bka} 
A.~Alloul, N.~D.~Christensen, C.~Degrande, C.~Duhr and B.~Fuks,
``FeynRules  2.0 - A complete toolbox for tree-level phenomenology,''
Comput.\ Phys.\ Commun.\  {\bf 185}, 2250 (2014)
doi:10.1016/j.cpc.2014.04.012
[arXiv:1310.1921 [hep-ph]].





\bibitem{Degrande:2011ua} 
C.~Degrande, C.~Duhr, B.~Fuks, D.~Grellscheid, O.~Mattelaer and T.~Reiter,
``UFO - The Universal FeynRules Output,''
Comput.\ Phys.\ Commun.\  {\bf 183}, 1201 (2012)
doi:10.1016/j.cpc.2012.01.022
[arXiv:1108.2040 [hep-ph]].



\bibitem{Alwall:2011uj} 
J.~Alwall, M.~Herquet, F.~Maltoni, O.~Mattelaer and T.~Stelzer,
``MadGraph 5 : Going Beyond,''
JHEP {\bf 1106}, 128 (2011)
doi:10.1007/JHEP06(2011)128
[arXiv:1106.0522 [hep-ph]].



\bibitem{Alwall:2014hca} 
J.~Alwall {\it et al.},
``The automated computation of tree-level and next-to-leading order differential cross sections, and their matching to parton shower simulations,''
JHEP {\bf 1407}, 079 (2014)
doi:10.1007/JHEP07(2014)079
[arXiv:1405.0301 [hep-ph]].





\bibitem{Pumplin:2002vw} 
J.~Pumplin, D.~R.~Stump, J.~Huston, H.~L.~Lai, P.~M.~Nadolsky and W.~K.~Tung,
``New generation of parton distributions with uncertainties from global QCD analysis,''
JHEP {\bf 0207}, 012 (2002)
doi:10.1088/1126-6708/2002/07/012
[hep-ph/0201195].




\bibitem{Jager:2010zm} 
B.~Jager,
``Next-to-leading order QCD corrections to Higgs production at a future lepton-proton collider,''
Phys.\ Rev.\ D {\bf 81}, 054018 (2010)
doi:10.1103/PhysRevD.81.054018
[arXiv:1001.3789 [hep-ph]].




\bibitem{Artoisenet:2012st} 
P.~Artoisenet, R.~Frederix, O.~Mattelaer and R.~Rietkerk,
``Automatic spin-entangled decays of heavy resonances in Monte Carlo simulations,''
JHEP {\bf 1303}, 015 (2013)
doi:10.1007/JHEP03(2013)015
[arXiv:1212.3460 [hep-ph]].




\bibitem{Frixione:2007zp} 
S.~Frixione, E.~Laenen, P.~Motylinski and B.~R.~Webber,
``Angular correlations of lepton pairs from vector boson and top quark decays in Monte Carlo simulations,''
JHEP {\bf 0704}, 081 (2007)
doi:10.1088/1126-6708/2007/04/081
[hep-ph/0702198 [HEP-PH]].



\bibitem{Sjostrand:2003wg} 
T.~Sjostrand, L.~Lonnblad, S.~Mrenna and P.~Z.~Skands,
``Pythia 6.3 physics and manual,''
hep-ph/0308153.



\bibitem{Sjostrand:2007gs} 
T.~Sjostrand, S.~Mrenna and P.~Z.~Skands,
``A Brief Introduction to PYTHIA 8.1,''
Comput.\ Phys.\ Commun.\  {\bf 178}, 852 (2008)
doi:10.1016/j.cpc.2008.01.036
[arXiv:0710.3820 [hep-ph]].




\bibitem{Cacciari:2011ma} 
M.~Cacciari, G.~P.~Salam and G.~Soyez,
``FastJet User Manual,''
Eur.\ Phys.\ J.\ C {\bf 72}, 1896 (2012)
doi:10.1140/epjc/s10052-012-1896-2
[arXiv:1111.6097 [hep-ph]].



\bibitem{Soyez:2008pq} 
G.~Soyez,
``The SISCone and anti-k(t) jet algorithms,''
doi:10.3360/dis.2008.178
arXiv:0807.0021 [hep-ph].




\bibitem{Contino:2016jqw} 
R.~Contino, A.~Falkowski, F.~Goertz, C.~Grojean and F.~Riva,
``On the Validity of the Effective Field Theory Approach to SM Precision Tests,''
JHEP {\bf 1607}, 144 (2016)
doi:10.1007/JHEP07(2016)144
[arXiv:1604.06444 [hep-ph]].




\bibitem{Englert:2014cva} 
C.~Englert and M.~Spannowsky,
``Effective Theories and Measurements at Colliders,''
Phys.\ Lett.\ B {\bf 740}, 8 (2015)
doi:10.1016/j.physletb.2014.11.035
[arXiv:1408.5147 [hep-ph]].




\bibitem{Farina:2016rws} 
M.~Farina, G.~Panico, D.~Pappadopulo, J.~T.~Ruderman, R.~Torre and A.~Wulzer,
``Energy helps accuracy: electroweak precision tests at hadron colliders,''
Phys.\ Lett.\ B {\bf 772}, 210 (2017)
doi:10.1016/j.physletb.2017.06.043
[arXiv:1609.08157 [hep-ph]].






\bibitem{Butterworth:2008iy} 
J.~M.~Butterworth, A.~R.~Davison, M.~Rubin and G.~P.~Salam,
``Jet substructure as a new Higgs search channel at the LHC,''
Phys.\ Rev.\ Lett.\  {\bf 100}, 242001 (2008)
doi:10.1103/PhysRevLett.100.242001
[arXiv:0802.2470 [hep-ph]].




\bibitem{Dokshitzer:1997in} 
Y.~L.~Dokshitzer, G.~D.~Leder, S.~Moretti and B.~R.~Webber,
``Better jet clustering algorithms,''
JHEP {\bf 9708}, 001 (1997)
doi:10.1088/1126-6708/1997/08/001
[hep-ph/9707323].



\bibitem{Wobisch:1998wt} 
M.~Wobisch and T.~Wengler,
``Hadronization corrections to jet cross-sections in deep inelastic scattering,''
In *Hamburg 1998/1999, Monte Carlo generators for HERA physics* 270-279
[hep-ph/9907280].

		
	
\end{thebibliography}
%


%

\end{document}